# THE ABUNDANCE OF DEUTERIUM
# IN THE WARM NEUTRAL MEDIUM
# OF THE LOWER GALACTIC HALO[1]

Blair D. Savage[2], Nicolas Lehner[3], Andrew Fox[4], Bart Wakker[2] and Kenneth Sembach[5]

## ABSTRACT

We use high resolution spectra obtained with the Far-Ultraviolet Spectroscopic Explorer (FUSE) and the Space Telescope Imaging Spectrograph (STIS) to obtain Milky Way interstellar column densities of H I, D I, O I, S II,  Fe II and P II toward the QSO HE 0226-4110 in the Galactic direction l = 253.4° and b = -65.77°. We obtain D/H = 21±(+8, -6) ppm from an analysis of the strong damped Lyman α line of H I and the weak higher Lyman series absorption of D I.  Correcting for a small amount of foreground contamination from D and H in the Local Bubble we obtain D/H = 22 (+8, -6) for the warm neutral medium of the lower Galactic halo.   The medium sampled has [O/H] = 0.12(+0.41, -0.20) and [Fe/H] = -1.01 (+0.10, -0.09). This suggests the abundances in the gas in the halo toward HE 0226-4110 are not affected by the infall of low metallicity gas and that the gas originates in the disk and is elevated into the halo by energetic processes which  erode but do not totally destroy the dust grains.  We compare our result to measured values of D/H in other astrophysical sites.  The value we measure in the halo gas is consistent with the hypothesis that for many Galactic disk lines of sight D is incorporated into dust.  The high average value of D/H = 23.1±2.4(1σ) ppm measured along 5 sight lines through disk gas in the Solar neighborhood is similar to D/H in the lower Galactic halo. These disk and halo observations imply the abundance of deuterium in the Galaxy has only been reduced by a factor of 1.12±0.13 since its formation.   In contrast, current galactic chemical evolution models predict the astration reduction factor should be in the range from 1.39 to 1.83. However, adjustments of uncertain parameters in these chemical evolution models may allow agreement to be achieved between the observed and predicted Galactic deuterium astration.

[2] Department of Astronomy, University of Wisconsin-Madison, 475 North Charter Street, Madison, WI 53706

[3] Department of Physics, University of Notre Dame, 225 Nieuwland Science Hall, Notre Dame, IN 46566

[4] Institut d'Astrophysique de Paris, 98bis Boulevard Arago, 75014 Paris, France

[5] Space Telescope Science Institute, Baltimore, MD 21218

*Subject heading:* cosmology: observations – Galaxy: abundances - ISM: abundances – ultraviolet: ISM

## 1. INTRODUCTION

Measures of the ratio of deuterium to hydrogen (D/H) in different Galactic and extragalactic environments can provide important insights about the nucleosynthetic history of those environments. Deuterium is believed to only form during the hot phases of the Big Bang and to be subsequently destroyed when it is converted to heavier elements in stars (Epstein, Lattimer & Schramm 1976). Consequently, as stars evolve and lose mass, they return to the interstellar medium (ISM) matter deficient in D. It is therefore expected that D/H in the ISM will decrease with time with the amount of the decrease (astration) providing information about the amount of cycling of matter through stars. A possibly serious observational complication regarding this simple picture could occur if the D in the ISM becomes selectively incorporated into interstellar dust as proposed by Jura (1982) and Draine (2004, 2006). In that situation measures of D/H in the gas phase can become confused by the combined effects of the destruction of D by astration and the selective removal of D from the gas by depletion.

To understand Galactic astration and to study the complications that are possibly produced by the depletion of deuterium into interstellar dust it is important to obtain measures of D/H in the gas phase in different physical and chemical environments that have not been substantially modified by chemical fractionation. There currently exist measures of D/H in the following locations: (1) In the low metallicity gas of the intergalactic medium (IGM) as recorded by QSO absorption line spectra (see O'Meara et al. 2001; Kirkman et al. 2003; Pettini 2006 and references therein). (2) In Complex C, a low metallicity high velocity cloud (HVC), falling into the Milky Way (Sembach et al. 2004). (3) In the Local Bubble extending to distances of ~100 pc from the Sun (see Moos et al. 2002 and references therein). (4) In the Solar neighborhood extending ~0.1 to 3 kpc from the Sun (see the recent summary papers of Linsky et al. 2006 and Hébrard 2006). (5) In a line of sight extending through the outer Galactic disk (Rogers et al. 2005). (6) In the atmosphere of Jupiter (Trauger et al. 1973; Mahaffy et al. 1998; Lellouch et al. 2001) and less accurately the atmosphere of Saturn (Lellouch et al. 2001; Griffin et al. 1996). In locations (1) to (4) D/H has been measured through observations of H I and D I Lyman series absorption lines. In the outer Galactic disk D I and H I hyperfine emission observations were employed. In the atmosphere of Jupiter three independent methods have been used to determine D/H. Note that we have omitted from this list environments where chemical fractionation effects are certain to be important in affecting the observed value of D/H (e.g. the Earth, Uranus, Neptune, and interstellar clouds where D is detected via its presence in HD and more complex molecules).

In this paper we report the first measurement of the value of D/H in the warm neutral medium (WNM) of the Milky Way halo along the line of sight to the bright QSO HE 0226-4110 ($z_{em} = 0.495$) in the Galactic direction $l = 253.94°$ and $b = -65.77°$. With a far-ultraviolet flux near 1000 Å of $2x10^{-14}$ erg cm$^{-1}$ s$^{-1}$ Å$^{-1}$, the QSO is bright enough to produce moderate signal-to-noise (S/N) high resolution spectra with the Far-Ultraviolet Spectroscopic Explorer (FUSE) satellite  and the Space Telescope Imaging Spectrograph (STIS) over the wavelength range from 920 to 1800 Å.  The resulting spectra clearly reveal the Lyman series absorption lines of H I and D I produced by the ISM.   The column density of H I is determined from the damping wings of the extremely strong Ly $\alpha$ line while the column density of D I is determined from the weak  Lyman series absorption lines of D I.   The QSO has a relatively smooth power-law continuum against which the absorption is measured.  Therefore, the continuum normalized Ly $\alpha$ absorption profile is well determined and the measurement of logN(H I) is reliable. The determination of D/H is also aided by the near absence of interstellar $H_2$ absorption which can produce serious ISM blending problems for lines of sight where the ISM $H_2$ column density is large.

The FUSE and STIS observations we have used to determine the value of D/H in the Galactic halo toward HE 0226-4110 have been used previously for a number of Galactic and extragalactic studies including:  (1) A study of the properties of Galactic HVCs  along the line of sight (Fox et al. 2005); (2) A study of Ne VIII in  the warm-hot intergalactic medium (Savage et al. 2005); (3)  A study of the QSO-associated absorber (Ganguly et al. 2006); and (4) A study of the low redshift intergalactic medium (Lehner et al. 2006).

Our paper is organized as follows:  In §2 we discuss the properties of the Galactic line of sight to HE 0226-4110. The FUSE and STIS observations and data processing methods are discussed in  §3.   The kinematic properties of the absorption are presented in  §4.  The techniques used to determine column densities for various atoms are discussed in §5.  The final abundance results for the line of sight to HE 0226-4110 are given in §6.  The implications of the values of D/H found in the Galactic halo and elsewhere are  discussed  in §7.

## 2. THE GALACTIC LINE OF SIGHT TO HE 0226-4110

The path to HE 0226-4110 (l = 253.94º and b = -65.77º) samples neutral Galactic gas in the Local Bubble, the low halo, and the distant halo.  In addition,  low column density positive high velocity absorption in H I and many other ions reveals gas  that is likely associated with the Magellanic Stream (Fox et al. 2005).

The 21 cm emission toward HE 0226-4110 obtained with the 34' beam Villa Elisa telescope ( Arnal et al. 2000) is shown in Figure 1 of Wakker et al. (2003).  The H I emission is well represented with two Gaussian components at Local Standard of Rest (LSR) velocities of  –7 and +12 km s$^{-1}$ with hydrogen column densities, N(H I) = $(1.55\pm0.01)x10^{20}$ and $(3.2\pm0.2)x10^{19}$ cm$^{-2}$, and profile widths (FWHM) of 17.9 and 15.4 km s$^{-1}$.  These line profile widths correspond to Doppler parameters b = FWHM/1.67 of 10.7 and 9.2 km s$^{-1}$, respectively.  Broad wings on the H I 21 cm emission profile appear to be artifacts of the baseline correction process as they are inconsistent with the ultraviolet absorption observations presented in this paper.

The effects of Galactic rotation along the high latitude line of sight to HE 0226-4110 are expected to be small.  Assuming the co-rotation of halo gas with gas in the underlying disk,  the radial velocity deviations due to Galactic rotation increase by ~+2 km s$^{-1}$ per kpc over the first 5 kpc along the line of sight.  Ignoring the effects of inflow or outflow the expected LSR velocity of halo gas absorption along the line of sight is near 0 km s$^{-1}$.

The value of the Galactic neutral hydrogen column density in the low velocity gas from an analysis of the Lyman series absorption in this paper is found to be logN(H I) = 20.12 (+0.04, -0.06) (see §5.5). FUSE measures of the logarithmic column density of molecular hydrogen in the low velocity gas are logN(J) = 13.65±0.21, 14.28±0.15, 13.82±0.20, 13.91±0.20, and ≤13.94, for the rotational quantum numbers J = 0, 1, 2, 3, and 4 respectively (Wakker 2006). The total logarithmic H$_2$ column density of log N(H$_2$) = log[∑N(J)] =14.58 in the J=0-3 levels is among the lowest recorded for any line of sight through the entire halo of the Galaxy. Note that Wakker (2006) has reported H$_2$ detections while Gillmon et al. (2006) instead report a 3σ limit of logN(H$_2$) ≤14.29. The small value of logN(H$_2$) is important for the determination of N(D I) and N(O I) because the effects of H$_2$ blending are greatly reduced for lines of sight with such low column densities of H$_2$.

The 0.25 keV X-ray emission in the general direction of HE 0226-4110 is relatively bright (Snowden et al. 1997) suggesting that the QSO lies in a direction that may pass through the southern opening of the Local Chimney (Welsh et al. 1999), which may be venting gas into the Galactic halo. This could explain why the column density of H$_2$ is very small in the direction to HE 0226-4110.

Although the distribution of the neutral gas as a function of distance has not been well studied toward HE 0226-4110, it appears that most of the H I absorption occurs beyond the Local Bubble at distances > 100 pc. We estimate the Local Bubble contribution to N(H I) in the direction to HE 0226-4110 from the Lehner et al. (2003) survey of absorption toward white dwarf stars ranging in distance from 49 to 200 pc. The survey includes three white dwarfs in the south Galactic polar direction with –65° < b < –84°. The distances range from 53 to 96 pc and average 79 pc. The values of logN(O I) toward these three stars range from 15.26 to 15.89 and average 15.54. Therefore, we would expect the first ~80 pc path through the Local Bubble toward HE 0226-4110 to produce logN(O I) ~ 15.54. Neutral oxygen is an excellent proxy for neutral hydrogen because both species have nearly the same ionization potential and they are strongly coupled to each other through charge exchange reactions. Therefore, using the measured value log[N(H I)/N(O I)] = 3.46 for gas in the Local Bubble from Oliveira et al. (2005), we estimate the Local Bubble contribution to the H I column density toward HE 0226-4110 to be logN(H I) ~ 19.00. The amount of neutral hydrogen associated with the first 80 pc of the Local Bubble in the direction to HE 0226-4110 therefore represents only ~8% of the total amount of neutral hydrogen at low velocity detected along the line of sight. We note that Sfeir et al. (1999) place the edge of the Local Bubble at logN(H I) = 19.3. With this value of N(H I) the Local Bubble contribution to the total amount of H I along the line of sight increases to 15%. Therefore, most of the low velocity Galactic H I in the direction of HE 0226-4110 likely exists in the thick disk of warm neutral gas extending into the Galactic halo that is estimated to have a scale height of ~ 360 pc (Dickey & Lockman 1990; Diplas & Savage 1994). Therefore, measures of D/H in the low velocity gas toward HE 0226-4110 will likely be dominated by absorption from gas in the extended WNM of the Galactic halo. The 17.9 km s$^{-1}$ width (FWHM) of the primary H I 21 cm emission component at –7 km s$^{-1}$ implies the warm medium has T <

6930 K; the actual temperature depends on the amount of the turbulent broadening contribution to the H I line width.

Although there are low column density intermediate velocity clouds (IVCs) near $-33$ km s$^{-1}$ (see §4), the absence of IVCs with $-100 < v < -60$ km s$^{-1}$ and HVCs with $v < -100$ km s$^{-1}$ is important since the D I absorption occurs at $-82$ km s$^{-1}$ in the H I rest frame. The absence of IVC absorption from $-60$ to $-100$ km s$^{-1}$ makes it possible to measure D I in the principal H I absorption component at $-7$ km s$^{-1}$ without confusion from negative velocity IVC and HVC H I absorption (see §5.4). Although there is low column density IVC absorption in the velocity range from $+20$ to $+40$ km s$^{-1}$, it does not affect the determination of reliable column densities for D and H in the principal absorption near $-7$ km s$^{-1}$.

HVC absorption is seen at velocities of 92, 146, 173, and 200 km s$^{-1}$ toward HE 0226-4110 in the lines of H I, C II, C III, C IV, O VI, Si II, Si III, and Si IV (Fox et al. 2005), even though no corresponding 21 cm H I emission is detected. These absorbing clouds appear to trace outlying fragments of the Magellanic Stream whose main 21 cm emission lies $\sim$11° away on the sky from HE 0226-4110. The body of the Magellanic Stream has 21 cm H I emission at velocities similar to those found in the absorption line HVCs toward HE 0226-4110. The positive velocity HVCs in the direction to HE 0226-4110 do not interfere with the analysis of the value of D/H in the principal H I absorption component at $-7$ km s$^{-1}$. Fortunately, the H I column densities in these HVCs are small enough that they do not affect the measurement of N(H I) derived from an analysis of the H I Ly $\alpha$ damping wings (see §5.5).

## 3. OBSERVATIONS AND DATA PROCESSING

The FUSE and STIS observations of HE 0226-4110 used in this study have been used for spectroscopic studies of HVCs (Fox et al. 2005), the IGM (Savage et al. 2005; Lehner et al. 2006) , and the gas in an associated absorption line system (Ganguly et al. 2006). Those papers provide very detailed discussions of the data handling procedures. Therefore the following discussions will be brief.

### 3.1. STIS Observations

The STIS observations were obtained between 2002 December and 2003 January in the 0.2"x0.06" aperture with the E140M Echelle grating which produces spectra with a resolution of 6.7 km s$^{-1}$(FWHM) from 1150-1700 Å with several small gaps between the orders at $\lambda > 1600$ Å. For details about the instrument and its in-orbit performance see Woodgate et al. (1998) and Kimble et al. (1998). The unbinned pixel size for the E140M grating is a function of wavelength and ranges from 3.0 to 3.2 km s$^{-1}$. The total integration time of 43.5 ks was achieved in 6 separate integrations which were aligned in wavelength and combined. The S/N per 7 km s$^{-1}$ resolution element is 11, 11, and 8 at 1250, 1500, and 1600Å, respectively. The S/N is substantially lower for $\lambda < 1250$ Å and for $\lambda > 1600$Å.

All velocities reported in this paper are in the LSR frame. Toward HE 0226-4110 the LSR and heliocentric velocity scales are related by $v_{LSR} = v_{helio} - 14.0$ km s$^{-1}$. The STIS reductions yield spectra with velocity uncertainties of $\sim$1 km s$^{-1}$ and occasional errors as large as 3 km s$^{-1}$ (see the appendix of Tripp et al. 2003). From the STIS spectra we measured the velocities of the blend free ISM lines of S II $\lambda\lambda$1250, 1253, 1259, N I

λλ1199, 1200, 1201, O I λ1302, Si II λλ1190, 1193, 1304, 1526, Fe II λ1608, Ni II λ1370, and Al II λ1670, and found $<v_{ISM}>_{LSR}$ = -3.6 ±1.4 km s$^{-1}$. This velocity for the principal ISM absorption was used to establish the zero point velocity calibration of the FUSE observations. Note that this velocity differs by ~3 km s$^{-1}$ from the velocity of –7 km s$^{-1}$ seen in the primary component of H I 21 cm emission (see §2).

### 3.2. *FUSE Observations*

The FUSE observations of HE 0226-4110 were obtained between 2000 December and 2003 January in the time tagged mode using the 30"x30" aperture (see Table 1 of Savage et al. 2005 for the exposure log). For information about FUSE and its in-orbit performance see Moos et al. (2000) and Sahnow et al. (2000). Four spectrograph optical channels (LiF1, LiF2, SiC1, and SiC2) are imaged onto two detectors which are divided into two independent segments (A and B). The total exposure times for HE 0226-4110 are 194 ks on detector 1 (segments 1A and 1B) and 191 ks on detector 2 (segments 2A and 2B). The night time observations, which are important for reducing airglow contamination, account for ~65% of the total exposure times. In the wavelength range from 916 to 987 Å we used only SiC2A measurements because of problems in aligning the individual faint SiC1B channel measurements. Over the wavelength range from 987 to 1187 Å we used observations from the LiF1A segment. From 1087 to 1182, LiF2A segment observations were used. Lower signal-to-noise observations in the LiF1B and LiF2B segments were used to check for fixed-pattern noise in the LiF1A and LiF2A segment observations. The final combined day+night integrations have S/N per 20 km s$^{-1}$ resolution element of 8, 17 and 17 at 950, 1000 and 1125 Å, respectively.

The FUSE wavelength calibration procedures (see Lehner et al. 2006) brought the FUSE observations into the same LSR reference frame as the STIS observations by forcing the velocities of the ISM lines of Si II, Ar I, Fe II and O I in the FUSE band to agree with the value $<v_{LSR}>$ = -3.6 km s$^{-1}$ observed for ISM lines in the STIS observations. The calibrated FUSE observations have a 1σ relative velocity error of ~5 km s$^{-1}$. The FUSE zero point velocity offset error is also estimated to be ~5 km s$^{-1}$.

FUSE spectroscopic observations are highly oversampled with 2 km s$^{-1}$ wide pixels. The spectra were rebinned to 4 pixels corresponding to 8 km s$^{-1}$ sampling. The re-sampling provides 2.5 samples per ~20 km s$^{-1}$ resolution element for the LiF channels and 3 samples per ~25 km s$^{-1}$ resolution element for the SiC channels.

### 4. KINEMATIC PROPERTIES OF THE ISM ABSORPTION

Continuum normalized interstellar absorption line profiles observed by STIS and FUSE for selected metal ions are shown in Figure 1. Basic absorption line measurements are given in Tables 1, 2 and 3. The errors are ±1σ and the limits are 3σ. The velocity structure of the absorption is summarized in Table 4. In Figure 1 the higher resolution STIS observations with 3.5 km s$^{-1}$ binning are easily distinguished from the FUSE observations with the coarser 8 km s$^{-1}$ binning. The higher resolution STIS observations are important for establishing the kinematic structure of the principal neutral gas absorption component near –4 km s$^{-1}$. The velocity structure of that absorption component is best determined from the absorption lines of the S II λλ1250, 1253, 1259 triplet, which mostly traces gas where the hydrogen is neutral. The principal absorption can be adequately modeled as a single component with b ~ 10 km s$^{-1}$ centered at $v_{LSR}$ ~ -4 km s$^{-1}$. Although the principal absorption component could contain substructure, such structure is not evident at the STIS resolution of 7 km s$^{-1}$. H I 21 cm emission line observations obtained with the Villa Elisa radio telescope (Arnal et al. 2000) reveal H I components at −7 and +12 km s$^{-1}$ with b = 10.7 and 9.2 km s$^{-1}$ and logN(HI) = 20.19 and 19.51, respectively. Spatially distinct regions smeared together in the 34' antenna beam of the Villa Elisa telescope are the likely reason the 21 cm HI emission reveals two components separated by 19 km s$^{-1}$ instead of a single component as seen along the infinitesimal UV absorption beam. The H I emission measurements show a simple kinematic structure for the principal H I emitting gas along the line of sight; the measurements are well fitted with a component having b(H I) = 10.7 km s$^{-1}$. This is important since the radio observations have a high spectral resolution (~1 km s$^{-1}$). The small velocity offset between the UV absorption at $v_{LSR}$ = -3.6±1.4 km s$^{-1}$ and the H I 21 cm emission component at $v_{LSR}$ = -7 km s$^{-1}$ may also result from beam sampling differences.

In the very strong absorption lines illustrated in Figure 1 there is evidence for other absorbing components at velocities well displaced from −4 km s$^{-1}$. The overall velocity structure of the absorption is summarized as follows:

*Principal absorption component.* Absorption in a single component occurs near −4 km s$^{-1}$ with a b value of ~10 km s$^{-1}$ measured in S II. This is the dominant absorbing component with logN(H I) = 20.12(+0.04,-0.06) (see §5.5). This absorption is detected in H I, D I, C II, C III, N I, O I, Si II, Si III, P II, S II, S III, Fe II, Ni II, and H$_2$.

*IVC1 and IVC2 absorption.* IVC absorption is seen in the strong lines of C II, C III, Si II, and Si III to extend from ~ −30 to −60 km s$^{-1}$. The absorption is most clearly traced by the STIS observations of C II λ1335 and Si III λ1206. It also appears in STIS observations of Si II λλ1260, 1193. The absorption appears as an asymmetric wing with decreasing absorption strength from −30 to −60 km s$^{-1}$. The absorption is also seen in the lower resolution FUSE observations of C II λ1036 and C III λ977. In the case of O I λ1302, absorption in this velocity range is only seen in a weak component near −33 km s$^{-1}$ situated on the negative velocity wing of the principal O I absorption. There is no evidence for additional absorption in the strong metal lines at $v_{LSR}$ < -60 km s$^{-1}$.

*IVC3 absorption.* The extremely strong lines of C II 1334, 1036, Si III 1206, and the higher H I Lyman series lines imply the presence of low column density IVC absorption in the velocity range from ~20 to ~40 km s$^{-1}$. We have not attempted to model this absorption since it does not affect the derived H I and D I column densities. The presence of this IVC absorption is not surprising. Essentially all lines of sight through ~ > 200 pc of halo gas in the vicinity of the Sun reveal numerous low column density IVC absorption features in the very strong metal lines. For examples of IVCs in UV observations of metal lines obtained at 3 and 10 km s$^{-1}$ resolution and high S/N toward bright stars in the low halo see Howk, Savage and Fabian (1999) and Savage & Sembach (1996).

*HVC1, HVC2, HVC3, and HVC4 absorption.* Positive HVC absorption is seen in components with $v_{LSR}$ = 92, 146, 173 and 200 km s$^{-1}$ in the higher Lyman lines and a variety of metal line species including C II, C III, C IV Si II, Si III, Si IV, and O VI. The approximate properties of the H I Lyman absorption as determined by Fox et al. (2005) are listed in Table 4. Since the HVC H I column densities are small, these high positive velocity absorption features do not interfere with our ability to measure column densities for D I or H I  in the principal absorbing component near $v_{LSR}$ = -4 km s$^{-1}$ along this line of sight.

## 5. COLUMN DENSITIES

A number of different techniques have been used to measure the basic properties of the UV absorption in the spectrum of HE 0226-4110 including Voigt profile fitting (PF) and standard curve of growth (COG) methods.  In addition, we occasionally use the apparent optical depth (AOD) method of Savage & Sembach (1991).  Using different analysis methods to determine absorption parameters including column densities is valuable because comparisons of the results provide important insights about systematic errors.

The basic absorption line measurements are given in Table 1 where we list in columns (1) the ion, (2) the rest wavelength, $\lambda$,  in Å, (3) the value of log($\lambda$f), (4) the absorption equivalent width, $W_\lambda$ in mÅ, (5)  the line strength, log$W_\lambda/\lambda$,  (6) identification of the observations analyzed as being  night-only, N, or  night+day , N+D, measurements, (7) the spectrograph channel, and (8) notes when appropriate.   The equivalent widths and uncertainties were determined following the procedures of Sembach & Savage (1992).  The adopted uncertainties are ±1σ.  These errors include the effects of statistical noise, fixed pattern noise for FUSE observations when observations for two or more channels exist,  and systematic uncertainties of the continuum placement and the absorption range over which the absorption lines were integrated.  The continuum levels were obtained by fitting low-order (<4) Legendre polynomials within 500 to 1000 km s$^{-1}$ of each absorption line.

To measure Doppler parameters and column densities we use both the PF  and COG methods.   The PF method used the Voigt component software of Fitzpatrick & Spitzer (1997) although for the analysis of the H I Ly α and Ly β absorption we also used the program VPFIT developed by R. Carswell and profile fit software developed by K. Sembach.   For the FUSE LiF and SiC bands the instrumental spread function was assumed to be a Gaussian with FWHM = 20 and 25 km s$^{-1}$, respectively.  For the STIS observations the spread function was adopted from Proffitt et al. (2002).

The COG method used the minimization and  $\chi^2$ error derivation approach outlined by Savage, Edgar & Diplas (1990).  The program solves for logN and b independently in estimating the errors since the $\chi^2$ contours in the logN-b plane are generally banana shaped. The COG method yields larger errors for logN and b than the PF method for several reasons. The equivalent widths used in the COG fitting contain continuum placement errors. In consrrast the PF method does not generally allow for continuum uncertainties unless different continuua are adopted as part of the fitting process.  Also,  the PF method places stronger constraints on the allowed range of b because the instrumentally blurred model profiles are directly compared to the observed profile widths  and the observed profiles must be aligned in velocity space when multiple lines are fitted simultaneously. When analyzing weak to moderately saturated absorption lines,  the PF errors are probably closer to the true errors particularly when different possible continua placements are included as part of the fitting process.

Atomic data (rest wavelengths, f-values, and damping constants) used in our analysis are taken from Morton (2003) except for Fe II and Ni II where we adopt the f-values from Howk et al. (2000) and Jenkins & Tripp (2006), respectively. The Ni II equivalent width measurement for the $\lambda 1317.217$ line is a factor of 1.95 larger than for the $\lambda 1370.132$ line. However, the similar f-values suggest the equivalent width difference should be a factor of 0.93 for unsaturated absorption. One of the Ni II lines is probably contaminated by blended absorption or another by blended emission. Since we cannot infer which line is affected, we have not tried to determine logN(Ni II) from these discrepant absorption line strengths.

Relatively weak absorption lines with $\log W_\lambda/\lambda < -4.3$ have been detected for D I, Fe II, P II, and S III. For these ions the derived column densities are only slightly influenced by saturation corrections. In the case of S II and O I the detected absorption lines range in strength from moderately strong to very strong with $\log W_\lambda/\lambda > -4.25$. Large saturation corrections are therefore required when determining N(S II) and especially N(O I). Although absorption by N I, Si II, Si III, C II, and C III has been detected, the available lines are so strong it is impossible to derive reliable column densities for the principal absorption component because of the very large line saturation. However, in the case of Si II, Si III, C II, and C III it is possible to determine useful information about the relatively weak IVC absorption from –30 to –60 km s$^{-1}$ which is valuable for determining the behavior in the wings of the H I absorption when analyzing the D I absorption. Velocities and column densities for the IVC absorption are listed in Table 3 based on the PF and AOD methods. The velocities $v_-$ and $v_+$, in that table indicate the AOD integration ranges.

The following discussions are organized according to the species being studied. We start with a discussion of the determination of metal line column densities in the principal absorber and then turn to the analysis of the D I and H I absorption. The metal line analysis provides valuable information about the kinematical properties of the principal absorption which is useful when analyzing the D I and H I absorption.

### 5.1. P II Absorption

The absorption line of P II $\lambda 1152$ listed in Table 1 is relatively weak with $\log W_\lambda/\lambda = -4.3$. Based on the well determined COG for Fe II discussed in §5.3 and shown in Figure 2, the P II line falls on the linear part of the COG assuming the b value for P II is similar to b(Fe II) = 10.5 ±0.9 km s$^{-1}$ derived from the Fe II COG. For P II we list a column density in Table 2 based on the AOD method. The final adopted column density and error are given in Table 5. Since the line saturation corrections are small, the dominant sources of error for this column density are the random error in the measurements and the continuum placement uncertainty.

### 5.2 The S II and S III Absorption

The absorption lines of the S II $\lambda\lambda 1250$, 1253, and 1259 triplet range from moderately strong ($\lambda 1250$) to very strong ($\lambda\lambda 1253$, 1259) (see Fig. 1 and Table 1). Therefore, the reliability of the derived S II column density is sensitive to the quality of the correction for line saturation. Column densities derived by the PF and COG methods are listed in Table 2. A single component Gaussian broadened COG applied to the measurements yields logN(S II) = 15.16 (+0.18,-0.12) and b(S II) = 10.8 (+2.8, -1.9) km s$^{-1}$. A single component Voigt profile fitted simultaneously to the triplet yields logN(S II) = 15.31±0.04 and b(S II) = 10.0±0.5. The 0.15 dex difference between the COG and PF result for logN(S II) illustrates the pitfalls of determining column densities from lines as strong as these.  In Figure 1 note that the simultaneous profile fit to the S II λ1250 line looks slightly too deep.  If this line is fit alone we instead obtain logN(S II) = 15.19±0.04 and b = 10.1±1.5 km s$^{-1}$.  This column density is 0.12 dex smaller than the simultaneous fit result and more consistent with the COG result.

For the final S II column density we adopt an average of the two  PF results, logN(S II) = 15.25±0.10.  The assigned 0.10 dex error spans the full range of ±1σ errors obtained from the two PF  results. Our final value for the Doppler parameter is b(S II) = 10.0 ±0.5 km s$^{-1}$, where we adopt the  smaller error from the simultaneous PF to the   S II triplet.

The S III λ1190.203 absorption illustrated in Figure 1b is clearly separated from the strong adjacent Si II λ1190.416 absorption.  The S III absorption is relatively broad and is likely well resolved in this STIS observation.  The PF value of logN(S III) = 14.56±0.09. The S III and S II absorption have similar values  of v but may differ in line widths,  b(S III) = 14.8±6.6 and b(S II) = 10.0±0.5 km s$^{-1}$,  both derived by the PF method.  Since S III mostly traces the warm ionized medium while S II mostly traces the WNM, the difference in line breadth is consistent with the observed trend of increasing line width  with increasing  ionization level found for lines of sight through the warm Galactic ISM (Sembach & Savage 1994).   The observed value of log[N(S II)/N(S III)] = -0.69,  provides insight about the degree of ionization of  the ISM in the halo toward HE 0226-4110. This is important when converting measures of N(S II) and N(H I) into estimates of  the S abundance, N(S)/N(H) (see §6.3).

### 5.3. The Fe II Absorption

The large number of observed transitions of Fe II span a range in line strength from weak to very strong (see Fig. 1 and Table 1).   The COG for the Fe II absorption is shown in Figure 2.  The COG shown is a simultaneous fit to all the measurements with b(Fe II) = 10.5 ±0.9 km s$^{-1}$.  The best fit column density is logN(Fe II) = 14.77±0.06. The error does not allow for the fact that the absorption may be somewhat more complex than a single Gaussian component.  However, the fact that four of the Fe II absorption lines lie on the linear part of the COG in the region with logW$_λ$/λ < -4.3 is a good indicator of the reliability of the column density estimate and its error.

A simultaneous PF to the Fe II absorption line observations with a single Gaussian component yields logN(Fe II) = 14.83±0.03 and b(Fe II) = 11.1±0.4 km s$^{-1}$. The PF error does not allow for continuum placement uncertainty in the individual Fe II profiles so the true error should be somewhat larger.

If we use the Fe II f-values recommended by Morton (2003) rather than Howk et al. (2002), the column density of Fe II from the COG and PF methods increases by only 0.01 dex.

The values of logN(Fe II) and b(Fe II) from the PF and COG method are in good agreement and the profile fits appear to reliably describe the observed profiles of both the strong and weak lines.  We adopt as the final column density logN(Fe II) = 14.80(+0.06,-0.09) with the ±1σ  errors chosen to  span the full range of errors determined by both the COG and PF methods.  Since many of the Fe II lines lie on the linear part of the COG, the resulting value of logN(Fe II) is not sensitive to the assumption that the actual optical depth  profile is well represented by a single component Gaussian.  We adopt the value

b(Fe II) = 10.8(+0.7, -1.2)  with the ±1σ error bars spanning the full range of b(Fe II) determined from the COG and PF methods.

## 5.4. The O I Absorption

To reduce the contamination of the O I absorption line measurements in the FUSE observations from airglow O I emission, we have analyzed the night-only observations. However, even those measurements are in some cases contaminated and we have omitted from the analysis leading to the estimate of logN(O I) the lines at λλ1039, 971, 948, and 936.  Blends with other absorbers have lead to the rejection of the  λλ950, 937 lines.

The value of logN(O I) derived from the FUSE and STIS observations is quite uncertain because the weakest detected O I lines without  airglow emission or blending problems including O I λλ930, 925, 922 (logW$_\lambda$/λ = -4.06,  –4.08 and –4.20) are relatively strong and  significantly affected by line saturation. For such lines small errors in the assumed kinematic model describing the absorption can introduce large systematic uncertainties in the derived value of logN(O I).

A single-component Doppler broadened COG fit to the O I lines free of airglow emission or blending (see Figure 2) yields logN(O I) = 16.95 (+0.37, -0.20) and b(O I) = 8.3±0.7 km s$^{-1}$.  A simultaneous single Voigt component PF to the same set of lines yields logN(O I) = 16.86±0.10 and b(O I) = 8.5±0.3 km s$^{-1}$.  The small change in the best fit value of b from 8.3 to 8.5 km s$^{-1}$ produces a substantial change in the derived best fit value of logN(O I).  We average the PF and COG results and adopt logN(O I) = 16.90 (+0.37, -0.20).   Since the absorption lines are so saturated, the derived column density is highly dependent on the assumed validity of the simple single-component kinematic model.  We therefore we adopt the large 1σ errors derived from the COG fit.

## 5.5. The D I Absorption

The value of logN(D I) has been derived from an analysis of absorption in the higher D I Lyman series absorption lines, including  λλ 949.48,  937.55,  930.50,  925.97, 922.90,  920.71, 919.10.  In Figure 3 we show the FUSE night-only absorption line measurements in the vicinity of the H I Ly λλ 972.54,  949.74,  937.80,  930.75,  926.23, 923.15,  920.96,  919.35, and 918.13 lines.  Hereafter we will refer to the D I and H I absorption lines using only the first three digits of the wavelengths listed above.  The measurements shown in Figure 3 are plotted in the H I reference frame and cover the velocity range from –300 to +300 km s$^{-1}$.   The left panel shows the observations before correcting for H I geocoronal emission. The right panel shows the observations after correcting for geocoronal emission.  The correction procedure is described below.  The D I absorption, which occurs at v$_{LSR}$ = -82 –7 =  –89 km s$^{-1}$ in the H I LSR reference frame, is highlighted in yellow for the detected lines of D I λλ949,  937,  930,  925  and the marginally detected lines of D I λλ922, 920.   We cannot rule out the possibility that some of the proposed D I absorption is from blue shifted H I. However, if this were the case the H I blue shift would need to be very similar to the –82 km s$^{-1}$ D I to H I isotope shift.

Even though our analysis has been restricted to orbital night-only observations, strong geocoronal H I emission severely contaminates the absorption in the D I λλ1025 and 972 lines. Weaker geocoronal contamination influences the D I λλ949, 937, 930 observations but is not a problem for the higher members of the D I Lyman series.

The H I geocoronal contamination to the D I observations was removed by first determining the profile shape of the emission from the strong emission seen in H I Ly λ972. The emission profile determined from this line was assumed to be the same for the higher  Lyman series geocoronal emission lines. The emission profile shape is produced by the full illumination of the FUSE 30"x30" aperture by the diffuse geocoronal emission and the shape of the profile should not significantly change from one Lyman line to the next. The smooth red profiles on the spectra for H I Ly λλ972, 949, 937, and 930 in the left panel of Figure 3 show our estimate of the geocoronal contamination affecting these Lyman series lines. The spectra illustrated in the right hand panel of Figure 3 show the FUSE spectra with the H I geocoronal emission removed. In the case of the Ly λ972 line, the removed emission is so strong that the resulting emission corrected measurements are very uncertain. However, for the H I λλ 949, 937, and 930  lines the geocoronal emission ranges from moderate to  weak in strength so the emission corrected observations should be reliable in the region of the D I absorption, particularly for the λλ937 and 930 lines.

The solid lines in the right panel of Figure 3 show the model H I absorption that affects the behavior of the continuum against which the D I absorption is observed. The H I absorption model shown in blue includes the primary component and the HVC components listed in Table 4. The primary component at $v_{LSR}$ = -4.1 km s$^{-1}$ produces the bulk of the absorption from –30 to + 30 km $^{-1}$ seen in all the H I Lyman series lines. The HVCs at 92, 146, 173 and 200 km s$^{-1}$ affect the absorption at positive velocities but do not influence the absorption near the D I lines. We have not attempted to model weak IVC absorption over the +20 to +40 km s$^{-1}$ velocity range. Ignoring this absorption has no effect on the derived D I  column density. A low column density H I component with logN(H I) = 16.67 and $v_{LSR}$ = -33.6 km s$^{-1}$ (as inferred from the  O I λ1302 absorption in IVC2) is required to fit the negative velocity wing of the H I absorption. The adopted b value of 14 km s$^{-1}$ for this absorber provides the best fit to the negative velocity wing of H I absorption. It is likely that this fitted component includes contributions from H I absorption associated with IVC1 detected in the ionized metal lines of C II, CIII, Si II, and Si III (see Table 3) and H I absorption associated with the O I absorption we identify as IVC2.   The FUSE 25 km s$^{-1}$ resolution does not allow a separation of H I absorption from IVC1 and IVC2.   The H I absorption model including the weak –33.6 km s$^{-1}$ component is shown with the red lines in the right hand panel of Figure 3. The following factors influenced the H I  fit process for this component: i) Forcing v(HI) = v(OI) in IVC2 at -33.6 km s$^{-1}$ is  the only physically reasonable assumption. ii) If [O/H] in IVC2 were < -3.45, then N(HI)>16.67, and  it would be necessary to lower b(HI) to fit the HI data in the range -80 to -40 km s$^{-1}$.  iii)  If [O/H] in IVC2 were > -3.45, then N(HI)<16.67, and it would be necessary  to raise b(HI) to fit the HI data in the range -80 to -40 km s$^{-1}$. iv) In either case,  the derived value of logN(D I) is unaffected, since whatever b(HI)/N(HI) combination is chosen, the continuum near DI has to be similar to the adopted continuum to have an acceptable fit.

The model describes well the rise of the flux with decreasing velocity on the negative velocity wings of the H I absorption (see Fig. 3). The wing fit is excellent for the H I λλ937, 926, 923, 920, 919, and 918 lines. For H I λ930 there appears to be a 5 km s$^{-1}$ shift which is consistent with the uncertainties in the FUSE wavelength scale. For H I $\lambda$949, uncertainties in the correction for the atmospheric H I emission make it difficult to access the reliability of the fit to the wing absorption.

The equivalent widths for the D I $\lambda\lambda$949, 937, 930, 925, 922, 920 and 919 absorption lines listed in Table 1 were derived using the continuum level established from the H I absorption model shown in Figure 3. The values of log [W$_\lambda$/$\lambda$] for these lines range from –4.84 to –4.12. With reference to the COGs for Fe II and O I, we see that the D I $\lambda\lambda$930, 925, 922, 920, and 919 lines have at most a modest amount of line saturation. Among these lines, the weak but reasonably well detected $\lambda\lambda$930 and 925 lines with logW$_\lambda$/$\lambda$ = -4.27 and -4.35 are very important in constraining the value of logN(D I).

We have measured the column density of D I in two ways. A simultaneous PF to the D I $\lambda\lambda$949, 937, 930, 925, 922, 920, and 919 measurements over the –70 to 30 km s$^{-1}$ velocity range yields logN(D I) = 15.45± 0.08, b(D I) = 10.4±2.1 km s$^{-1}$, and v$_{LSR}$ = -7.5±1.3 km s$^{-1}$. With a reduced $\chi^2$ =1.17, the fit is quite acceptable. The fit results are shown in Figure 4. The small velocity offset of the D I absorption from the velocity of the H I absorption of –4.1 km s$^{-1}$ is well within the range of uncertainty of the FUSE velocity calibration. A fit of the measured equivalent widths for the same lines to a simple one component Doppler broadened COG yields logN(D I) = 15.33 (+0.24, -0.19), b = 8.6 (+5.9, -2.4) km s$^{-1}$. The COG fit is shown in Figure 2.

The COG errors for logN(D I) and b are several times larger than the corresponding PF errors. The COG errors for logN(D I) are relatively large because the COG fit only poorly constrains the value of b (D I) = 8.6 (+5.9, -2.4) km s$^{-1}$. In contrast the simultaneous PF which directly evaluates the observed line widths and velocities yields b (D I) = 10.4±2.1 km s$^{-1}$. We would expect b(D I) to lie somewhere between b(O I) and b(H I). We find b(O I) = 8.3±0.7 and 8.5±0.3 km s$^{-1}$ from the COG and PF analysis applied to the O I absorption. We find b(H I) = 10.7 km s$^{-1}$ from the 21 cm emission observations (see §2). Therefore, b(D I) should be between 8.4 and 10.7 km s$^{-1}$. The b value constraints for D I determined from the PF are consistent with our knowledge of the b values of O I and H I. We therefore believe the PF result and its errors are more reliable than the COG result. Note that the value of b(D I) determined from the PF is consistent with the absorption being produced by D I. If the absorption were instead produced by H I absorption blue shifted by –82 km s$^{-1}$ from the primary H I absorption, there is no reason to expect the observed b value to be consistent with the properties of H I and O I absorption in the primary absorber.

Another reason why the COG errors are larger than the PF errors is that the values of the equivalent widths used in the COG analysis do allow for continuum placement uncertainty while the PF result in Table 2 does not allow for continuum placement uncertainty. To probe the sensitivity of the PF results to different choices of the continuum we repeated the PF for alternate but acceptable continuum levels and found values of logN(D I) within +0.03 and –0.05 dex of the best fit result listed above. Note continuum adjustments explored approximately allow for uncertainties associated with the model profile fits to the H I IVC absorption wing described by the red line in Figure 4. Note that the fitted b-value of the D I absorption is strongly constrained by the negative velocity side of the D I lines (i.e. the observations from -40 to 0 km/s in Fig 4), totally independent of the fit to the H I IVC2, which only affects the positive velocity side of the D I line.

The PF results are not sensitive to the choice of lines included in the fit process. Values of logN(D I) obtained by excluding λ919 or λ919, 920 or λ919, 920, 922 or λ919, 920, 922, 949, or λ919, 920, 922, 930, 949 all lie within ±0.03 dex of the best fit result given above. However, the smallest reduced $\chi^2$ of 1.17 is found when all the transitions from λ919 to λ949 are included.

The PF results are also not sensitive to adjustments in the D I absorption velocity. The best fit velocity obtained from the simultaneous D I profile fit to the full set of D I lines is −7.5±1.3 km s$^{-1}$. Fixing the fit velocity at −10.5 km s$^{-1}$ resulted in no change in derived value of the D I column density. Fixing the fit velocity at −4.5 km s$^{-1}$ yielded a 0.04 dex decrease in the derived value of the D I column density. Note that the errors included in the simultaneous profile fit already include the effects of the ±1.3 km s$^{-1}$ PF error in the D I velocity.

We probed the sensitivity of the derived column density to changes in the assumed width of the instrumental line profile. Increasing the width from 25 to 27 km s$^{-1}$ resulted in a 0.03 dex increase in the best fit D I column density. Decreasing the width from 25 to 23 km s$^{-1}$ produced a 0.03 dex reduction in the D I column density.

We adopt for the final column density and errors log N(D I) = 15.45 ±0.08 (+0.04, -0.06) where the first error is the PF statistical error and the second set of errors is the quadrature addition of the systematic errors associated with the continuum fit and the instrumental profile width. These errors represent the ±68% confidence range.

The derived value of logN(D I) for the line of sight to HE 0226-4110 is not seriously affected by saturation problems since four of the absorption lines contributing to the final result are only slightly influenced by line saturation effects. The dominant source of error in the reported value of the column density is the statistical uncertainty in the observations.

### 5.6. H I Absorption

The value of logN(H I) derived from profile fits to the extremely strong Ly α and Ly β absorption is primarily determined by the very broad damping wings on Ly α. The Ly α line is so strong that the behavior of the absorption in the line wings is completely dominated by the natural damping associated with the principal absorption component near −4 km s$^{-1}$.

We checked to see if the behavior of the QSO continuum in the vicinity of the Ly α and Ly β lines could be influenced by the presence of emission lines in the QSO spectrum. Knowing about these emission lines is important for judging the behavior of the continuum against which the H I ISM absorption is observed. Scott et al. (2004) present a composite EUV spectrum of QSOs derived from FUSE observations of AGNs (see their Figure 11). In the EUV rest-frame, QSOs have relatively smooth power law continua with superposed broad emission lines. The redshifted EUV spectral region of HE 0226-4110 illuminating the ISM Ly β 1025.72Å line has a rest wavelength of ~686 Å. The composite QSO spectrum is continuous over the wavelength region from 630 to 710 Å, suggesting that the QSO continuum illuminating the ISM Ly β absorption should be relatively smooth. In contrast, the redshifted EUV spectral region of HE 0226-4110 illuminating the ISM Ly α 1215.67Å line has a rest wavelength of ~813 Å. In QSOs that spectral region is bracketed by emission from Ne VIII λλ770, 780 and O IV λ789 which typically extends from 750 to 795Å and peaks near 772 Å, and weak emission by O III λ833 which extends from ~820 to 840Å and peaks near 833Å. The redshifted emission lines shortward and longward in wavelength of the ISM Ly α absorption need to be considered when fitting the continuum in the vicinity of the broad Ly α absorption. Most of the AGN emission lines found by Scott et al. (2004) over the rest frame wavelength region from 630 to 1160 Å are evident in the FUSE and STIS spectra of HE 0226-4110 shown by Lehner et al. (2006) and Ganguly et al. (2006).

The HE 0226-4110 Ly α absorption is displayed over a large (±10,000 km s$^{-1}$) and small (±2500 km s$^{-1}$) velocity range in the top three panels of Figure 5. For Ly β the measurements over the ±2500 km s$^{-1}$ velocity range are shown in the bottom panel of Figure 5. The continua we have adopted are shown as the solid green line placed on each spectrum. Regions of obvious emission or absorption were avoided when determining the continua. QSO emission from Ne VIII λλ770, 780 and O IV λ787 is seen in the velocity ranges from −10,000 to −7000 km s$^{-1}$ in the upper panel of Figure 5. QSO emission from O III λ833 is relatively weak but explains the slight rise in flux compared to the continuum in the velocity range from 5000 to 8000 km s$^{-1}$. For Ly α the continuum was determined from the observed flux in the velocity ranges from −7000 to −3000 km s$^{-1}$ and from 9500 to 21,000 km s$^{-1}$. These velocity ranges avoid those regions of the QSO spectrum influenced by Ne VIII+O IV emission and O III emission. A first order polynomial provides a good fit to the continuum although we tried higher order polynomials to evaluate uncertainties in the continuum fit and the effect of those uncertainties on the derived value of logN(H I).

Strong H I geocoronal emission is evident at the center of the Ly α and β lines. The Ly β line profile is also affected by strong HV H I absorption from +70 to +220 km s$^{-1}$, weak ISM O I absorption at 1025.76 Å, and by O I airglow emission at 1025.76, 1027.43 and 1028.16 Å. The O I emission occurs at +12, +500 and +713 km s$^{-1}$ in the Ly β reference frame. The O I emission at 500 and 713 km s$^{-1}$ is evident in the spectrum. The H I and O I atmospheric emission recorded by FUSE is very broad (FWHM ~100 km s$^{-1}$) because the diffuse emission fills the 30"x30" entrance aperture.

The kinematic model used to model the H I absorption is given in Table 4. The primary H I absorption component is set to $v_{LSR}$ = -4.1 km s$^{-1}$, the velocity of S II absorption obtained by a PF of the S II triplet observed in the same STIS spectrum as the Ly α line. The amount of H I in the primary absorption component exceeds that in the IV and HV components by more than a factor of $10^3$. Therefore, the presence of these additional components has no effect on the derived value of logN(H I) in the primary component. The IV and HV components produce absorption that is lost in the core of the Ly α line that has essentially zero intensity over a velocity range of ±300 km s$^{-1}$. In addition, the IV and HV components do not have large enough column densities to produce contributions to the damping wings in Ly α observed at large positive and negative velocities. The value of logN(H I) obtained from profile fits to the Ly α line is primarily determined by the properties of the absorption over the velocity range from 300 to 2500 km s$^{-1}$ and from -300 to -2500 km s$^{-1}$.

A Voigt profile fit to the Ly α absorption using the linear continuum shown in Figure 5 yields logN(H I) = 20.12±0.02, where the ±1σ error only includes the effects of random statistical noise. The Ly $\alpha$ fit is displayed in Figure 5 on the STIS observations with 7 km s$^{-1}$ (2 pixel) binning (second panel from the top) and with 52.5 km s$^{-1}$ (15 pixel) binning (third panel from the top). In the panel with 52.5 km s$^{-1}$ binning we also display absorption profiles for logN(H I) 2$\sigma$ = 0.04 dex higher and lower than the best fit value. A value of b(H I) = 10.7 km s$^{-1}$ was assumed for the fit based on the 21 cm emission profile. However, the damping wings on the Ly $\alpha$ line are so broad that increasing or decreasing b by factors of 4 has no effect on the derived value of logN(H I). Using logN(H I) = 20.12 and b = 10.7 km s$^{-1}$ along with the IVC and HVC components listed in Table 4, we obtain the Ly $\beta$ profile shown in the lower panel of Figure 5. The noiser Ly $\beta$ observations are consistent with the value of logN(H I) obtained from the Ly $\alpha$ line. Given the contamination of the Ly $\beta$ absorption by H I and O I air glow emission and by H I absorption in the HVCs, the Ly$\beta$ observations are not useful in constraining the size of the error for the derived value of logN(H I). That error is entirely determined from the quality of the fit to the Ly $\alpha$ line wings.

To perform the Voigt profile fitting to the Ly $\alpha$ line we used the VPFIT software package (developed by Robert Carswell) and used an independent set of fitting software. Both codes gave the same results and errors when applied to observations fitted with the same continuua.

A value of logN(H I) = 20.12±0.02 in the principal component toward HE 0226-4110 was obtained from a linear fit to the continuum measured near Ly $\alpha$ but avoiding the wavelength regions contaminated by QSO emission lines. In order to evaluate how uncertainties in the fitted continuum affect the derived value of logN(H I) we also obtained fits to the continuum using second, third, and fourth order polynomials. The fits using second and fourth order polynomials both yielded logN(H I) = 20.12±0.02. However, the fit using the third order polynomial produced a lower continuum in the vicinity of the Ly $\alpha$ line and the profile fit yielded logN(H I)= 20.08±0.02. While we believe the linear continuum fit to be the valid representation of the continuum that applies in the vicinity of the Ly $\alpha$ line we cannot rule out the lower continuum fitted with the third order polynomial. We take the -0.04 dex difference between the two derived values of logN(H I) to be a measure of the negative systematic error associated with uncertainty in the continuum fit. Therefore logN(H ) = 20.12±0.02 (+0.00, -0.04). These errors represent the ±68% confidence limits.

## 6. RESULTS

In this section we combine the column density measurements to determine total elemental abundances with respect to hydrogen. Measures of log[N(DI)/N(H I)] and log[N(O I)/N(H I)] are good measures of log[N(D)/N(H)] and log[N(O)/N(H)] in the gas since D I, O I, and H I exist in the same neutral regions, and D I and O I are absent in regions where the hydrogen is ionized. That is because D I, O I, and H I have essentially the same ionization potentials, and O I is very strongly coupled to H I via charge exchange reactions (see Sofia & Jenkins 1998). The molecular hydrogen fraction along the line of sight is extremely small: log[N(H$_2$)/N(H I)]= -5.49. Therefore, we do not need to worry about molecular fractionation effects involving HD and H$_2$. The issue of whether some of the deuterium along the line of sight may be incorporated into dust (including PAHs) is addressed in §7.

Fe II,  S II,  and P II are the dominant states of ionization in regions where the hydrogen is mostly neutral.  However, for lines of sight through the halo where an appreciable fraction of the H I is ionized,  substantial amounts of these singly ionized species may exist in the ionized regions (Howk et al. 2006).  Therefore, measures of log[N(X II )/N(H I)] are not necessarily reliable measures of log[N(X)/N(H)] in the gas phase.  For example, toward the halo star vZ 1128 in the globular cluster M3 [l = 42.2, b = +78.7, z = 10.2 kpc, and z = 10.0 kpc], Howk et al. (2006) estimate that N(H II)/N(H) ~ 0.43.  Hence,  an appreciable amount of the singly ionized ions may exist in the ionized gas.  Howk et al. (2006) determine for the line of sight to vZ1128 that   log[N(S)/N(H)] = log[N(S II)/N(H I)] - 0.17 and log[N(Fe)/N(H)] = log[N(Fe II)/N(H I)] - 0.07.  S II and Fe II existing in  regions of ionized hydrogen introduce the 0.17 and 0.07 dex  ionization corrections for S II and Fe II in the expressions above.  Since the path through halo gas to HE 0226-4110 may be similar to the path to vZ 1128,  we need to allow for possible ionization corrections when determining elemental abundances from the observations of column densities of the singly ionized species.  Fortunately, this problem does not affect our determination of log[N(D)/N(H)] or log[N(O)/N(H)].  However, the problem does affect the derived abundances for Fe, S and P.

### 6.1. *D/H toward HE 0226-4110*

Ionization corrections for D I or H I or corrections for $H_2$ are not important when determining D/H  for the line of sight to HE 0226-4110.  Using the final values of logN(D I) = 15.45 ±0.08 (+0.04, -0.06) and logN(H I) = 20.12 ±0.02 (+0.00, -0.04)  from Table 5, we obtain log [N(D I)/N(H I)] = -4.67 ±0.08±0.06  where  the random and systematic errors have separately been combined quadratically.  We combine the total random and total systematic errors linearly to obtain log [N(D I)/N(H I)] = -4.67 ±0.14.  Therefore, the value of D/H in the medium toward HE 0226-4110 is D/H = 21 (+8,- 6) $\times10^{-6}$ = 21 (+8, -6) ppm.  This value of D/H includes contributions from D and H in the Local Bubble    and the WNM of the Galactic halo, with most of the absorption occurring in the halo.  If we correct for the foreground Local Bubble absorption using logN(H I)$_{Local Bubble}$ = 19.0 (see §2) and <D/H>$_{Local Bubble}$ = 15.8±2.1 ppm  from Linsky et al. (2006), we obtain the slightly larger value (D/H) = 22 (+8, -6)  for the WNM of the halo. Within the two digit round off,  this value does not change if we instead adopt logN(H I)$_{Local Bubble}$ = 19.3 for the Local Bubble foreground contamination.

### 6.2. *D/O and O/H  toward HE 0226-4110*

Ionization corrections are not important in determining D/O and O/H from measurements of D I, O I, and H I in the WNM toward HE 0226-4110. We obtain from the final column densities listed in Table 5  log[N(O I)/N(H I)] = log(O/H) = -3.22 (+0.41, -0.20) and log[N(D I)/N(O I)] = log (D/O) = -1.45 (+0.26, -0.44).  The total errors listed here were obtained by first separately combining the random and systematic errors quadratically. The total single error listed is obtained from a linear combination of the total random and total systematic errors.  Although the errors are large, the result for log(O/H) is similar to the value of log(O/H)$_O$ = -3.34 for the Solar abundance of oxygen (Asplund et al. 2004).   Therefore, the oxygen abundance in the WNM toward HE 0226-410 is approximately Solar with [O/H] = log(O/H)-log(O/H)$_O$  = 0.12 (+0.41, -0.20).  This is consistent with the idea that the WNM in the low halo represents predominantly thin disk gas elevated into the halo by energetic processes occurring in the disk. For example, single and multiple supernova explosions can heat and over-pressurize disk gas which can then drive a circulation of gas between the thin and thick disks of the Galaxy. In such a 'fountain flow' one would expect the thick disk of the Galaxy to have elemental abundances resembling those found in the thin disk.

### 6.3. *S/H and Fe/H toward HE 0226-4110*

Since S and Fe have been detected in the singly ionized state, we need to worry about how to allow for ionization effects when converting $\log N(X\ II)/N(H\ I)$ to log $N(X)/N(H)$. An ionization correction is required since substantial quantities of these singly ionized atoms many exist in regions where the hydrogen is ionized. Therefore, the observed values of $\log N(X\ II)/N(H\ I)$ will be larger than log $N(X)/N(H)$.

We can gain some insight about the level of ionization along the line of sight to HE 0226-4110 since we have measures of column densities for S II and S III. We find that $N(S\ III)/N(S\ II) = 0.20$. This can be compared to the value toward vZ 1128 where $N(S\ III)/N(S\ II) = 0.15$. The ISM toward HE 0226-4110 contains ~33% more doubly ionized gas than the ISM toward vZ 1128 studied by Howk et al. (2006). As a rough approximation, we expect the ionization corrections for S II and Fe II to be ~33% larger for the HE 0226-4110 line of sight than for the carefully analyzed vZ 1128 line of sight. This would imply gas phase S and Fe abundances of

$\log[N(S)/N(H)] \cong \log[N(S\ II)/N(H\ I)] - 0.29 = -5.16\ (+0.14, -0.10)$

and

$\log[N(Fe)/N(H)] \cong \log[N(Fe\ II)/N(H\ I)] - 0.19 = -5.51(+0.10, -0.09)$

Using the Solar S and Fe abundances of $\log(X/H)_{O} = -4.80$ and $-4.50$, respectively, from Grevesse and Sauval (1998), we infer [S/H] = -0.36 (+0.14, -0.10) and [Fe/H] = -1.01 (+0.10, -0.09), where the errors do not allow for the uncertainty in the ionization correction which could be as large as ±0.20 dex. Note that the value of [S/H] decreases by 0.13 dex if we use the Solar photospheric abundance of S from Grevesse and Sauval (1998) rather than the meteoric value.

Since sulfur is not normally incorporated into interstellar dust (Savage & Sembach 1996), the value [S/H] = -0.36 ±(+0.14, -0.10) suggests a subsolar abundance of sulfur in the medium toward HE 0226-4110. In contrast we found [O/H] = +0.12(+0.41, -0.20) implying an approximately Solar oxygen abundance. The sulfur abundance estimate requires a large and uncertain -0.29 dex correction for the presence of S II in the ionized hydrogen along the line of sight. Without applying that ionization correction we obtain [S/H] = -0.07±(+0.14, -0.10). We believe the abundance estimate for oxygen, which does not require an ionization correction, provides the best estimate of the true metallicity of the gas along the line of sight.

The ~1.0 dex deficiency of Fe is almost certainly due to the incorporation of Fe into grains in the WNM of the halo. The modest level of depletion of Fe is consistent with the idea that substantial but not total dust destruction occurs in the explosive processes that elevate disk gas into the halo.

### 7. DISCUSSION

Measuring values of D/H in different astrophysical sites with varied nucleosynthetic histories and physical/chemical states is crucial for ultimately understanding the combined effects of astration and other possible variables on the deuterium gas phase abundance.

### 7.1. Observed values of D/H in different Astrophysical Sites

In Figure 6 we present a summary of measures of D/H according to the particular site of the measurement from near to far. The sources of the results presented in Figure 6 are summarized in Table 6 with additional comments below.

*Jupiter and Saturn*. D/H in the atmospheres of Jupiter and Saturn may provide good measures of the value of D/H in the Galaxy at the Solar radius $4.6 \times 10^9$ years ago or after ~8 billion years of the evolution of the Milky Way. This is because the main reservoir of deuterium in these major planets is in molecular hydrogen. A complication for both planets is the possible enrichment of deuterium by 5-10% and 25-40% due to the mixing of the protosolar nebular medium with deuterium enriched ices during the formation of the planets (Guillot 1999). Another possible complication is the complex physical and chemical behavior of the layering of the different phases of $H_2$ with depth into the planets (gaseous, liquid, liquid metallic) could cause the different layers to have different values of D/H. However, we are not aware of a careful theoretical discussion of this problem.

Three different techniques have been used to measure D/H in the atmosphere of Jupiter. Trauger et al. (1973) obtained D/H = 21± 4 ppm from observations of weak optical $H_2$ and HD absorption lines seen in the reflected sunlight. That result is supported by the less accurate observations of another HD absorption line by Smith et al. (1989) yielding D/H = 19.5±9.5. Mahaffy et al. (1998) obtained D/H = 26±7 ppm from Galileo probe mass spectrometer observations. Lellouch et al. (2001) obtained D/H = 24±4 ppm from Infrared Space Observatory observation of $H_2$ and HD infrared emission. An error weighted average of these values from Tauger et al., Mahaffy et al., and Lellouch et al. gives <D/H >= 23.0±2.6 (1σ). This value may be 5-10% larger than the protosolar nebular gas value if Jupiter received deuterium enriched ices during its formation (Guilott 1999).

D/H has also been determined for the atmosphere of Saturn from IR measures of HD and $H_2$ absorption and emission lines. Lellouch et al. (2001) and Griffin et al. (1996) report D/H = 18.5 (+8.5, -6) ppm and 23 (+12, -8) ppm, respectively. The errors on these measurements are large compared to those for Jupiter. In addition, Guilott (1999) estimates that a large 20-40% downward correction may be needed to determine the protosolar nebular value of D/H from these measurements because Saturn may have received a substantial enrichment of D during formation from deuterium enriched ices. Because of these complications we concentrate on the implication of the measurements for Jupiter hereafter.

*Local Bubble*. Extensive measures of D/H in the Local Bubble have been obtained by the FUSE and HST satellites (see Moos et al. 2002). Taking the Local Bubble values compiled by Linsky et al. (2006) we obtain a straight average and standard deviation for the direction to 22 stars of <D/H> = 15.8±2.1 (SD) ppm. The nearby stars range in distance from 3 to 100 pc and have logN(H I) <19.2. The small dispersion in the values for Local Bubble lines of sight implies a standard value of D/H in the gaseous medium close to the Sun.

Oliveira et al. (2006) have argued that the value of D/H in the Local Bubble is the same as found in the WNM of the Galactic disk where the average line of sight hydrogen density <n(H I)> is ~0.1 cm$^{-3}$.

*Solar Neighborhood.* We define the Solar neighborhood as being the region surrounding the Sun beyond the Local Bubble with logN(H I) extending from 19.3 to 20.7. This corresponds to distances from ~100 to ~1000 pc. The values of D/H toward 19 stars measured in this region (compiled by Linsky et al. 2006) span a large range with 12 values of D/H clustering at low values, 2 giving intermediate values, and 5 giving large values of D/H. In Figure 6 and Table 6 we illustrate this behavior of D/H by determining the average of the 12 low values and the average of the 5 high values. The low value of <D/H> =10.1±3.3 (SD) is the average and standard deviation for 12 lines of sight with D/H < <D/H>$_{Local Bubble}$. The high value of <D/H> = 21.7 ±1.7 ppm is the error weighted average and 1σ error for five lines of sight. Note that these two averages have not been corrected for a small amount of foreground contamination from H I and D I absorption in the Local Bubble. When that correction is applied to the high value using <D/H>$_{Local Bubble}$ = 15.8±2.1 (SD), the result increases to <D/H> = 23.1±2.4 ppm (Linsky et al 2006). This corrected value should be used when considering the Galactic nucleosynthetic implications of the measurement (see §7.3).

*Galactic Disk.* The Linsky et al. (2006) compilation of D/H values includes five Galactic disk lines of sight with logN(HI) > 20.7. The average and standard deviation for these observations are <D/H> = 8.5±1.0 (SD) ppm. At large distances through the Galactic disk where the observed values of logN(H) are large, the observed values of D/H are small. A possible exception is the high value of D/H = 21.4 (+5.1-4.3) measured toward HD 41161 by Oliveira & Hébrard (2006). However, it is difficult to evaluate the quality of this measurement since very few details are given in the paper about how line saturation might affect the derived value of logN(D I).

*Outer Galactic Disk.* Rodgers et al. (2005) have reported the secure detection of the weak D I spin flip hyperfine transition at 327 MHz (92 cm) in the direction of the Galactic anticenter with l = 183°. In this direction the 21 cm H I emission is extremely strong and affected by self absorption. Assuming a uniform spin temperature of 130 K and the uniform mixing of Galactic continuum emission along the line of sight they obtain D/H = 23±4 ppm. However, the small ±1σ error does not include the effects of the large systematic uncertainty associated with the analysis of the radiation transfer in the H I emission. Rodgers et al. (2005) present 3σ errors including the systematic effects of changing the H I spin temperature over the range from 110 to 150 km s$^{-1}$ and report D/H = 23 (+15, -13) ppm. However, the assumed range of spin temperature may be smaller than the actual range and the authors have not included the large systematic uncertainty associated with assumptions regarding the relative distributions of the continuum emission and line emission along the line of sight. A realistic systematic error associated with this observation is likely to be very large, so we have not plotted this result in Figure 6.

*Galactic Halo.* The single measurement for a line of sight through the WNM of the Galactic halo of D/H = 22 ±(+8, -6) ppm is from this paper for the direction to HE 0226-4110. This result has been corrected by +1 ppm due to the foreground contamination from H I and D I absorption in the Local Bubble. The approximately solar abundance for O for this line of sight suggests that halo WMN is populated by gas expelled from the thin Galactic disk. The observed Fe depletion (~1 dex) and solar abundance of O are typical of warm clouds in the Galactic halo (see Spitzer & Fitzpatrick 1993; Sembach & Savage 1996; Savage & Sembach 1996). This strongly suggests that the cloud(s) encountered along the HE 0226-4110 sight line near –4 km s$^{-1}$ are located at distances of several hundred pc or more away from the Galactic plane, and rules out the hypothesis that the absorption could be occurring more locally in cool diffuse clouds where the Fe depletion would be expected to be much larger. The observed Fe depletion is consistent with the idea that the injection of matter into the halo by supernovae occurring in the Galactic disk does not completely destroy the resistant cores of the grains. Although very little information is available about the possible existence of PAHs in the Milky Way halo, Lehner & Howk (2006) have reported the detection of PAH emission from regions 2-4 kpc above the mid-planes of several edge-on galaxies they have observed with the Spitzer IR observatory. Therefore, we cannot rule out the possibility that PAHs may exist in the halo gas and dust along the HE 0226-4110 line of sight. However, the extremely low abundance of $H_2$, with $N(H_2)/N(H\,I) = 2.9 \times 10^{-6}$ (Wakker 2006) implies the sight line is hostile to the formation and survival of simple molecules.

*Complex C*. Complex C is a low metallicity (~1/6 Solar) high velocity cloud falling into the Milky Way. For a discussion of the properties of Complex C see Wakker (2001). Complex C has a large angular extent (~1700 deg$^2$) and evidently contains no dust, $H_2$, or stars. Complex C is suspected to be either intergalactic gas or material stripped from a nearby galaxy falling into the Milky Way. From FUSE observations, Sembach et al. (2004) report D/H = 22±7 ppm in Complex C along the path to the QSO PG 1259+593. This observation provides information about the single site in the local Universe beyond the Galactic disk and halo where D/H has been measured.

*IGM*. In the distant Universe observations of D I and H I absorption in damped Lyman α systems and in Lyman limit systems have provided reliable values of D/H in 6 absorbing systems with relatively simple velocity structures. For a recent review of these and other less reliable results see Pettini (2006). The basic measurements for the simple systems can be found in the observational papers of O'Meara et al. (2001, 2006), Kirkman et al. (2003), and Pettini & Bowen (2001). In Figure 6 we plot D/H for the 6 individual measurements. The spread of values is considerably larger than would be expected based on the assigned observational errors assuming D/H should be the same in each absorber. In fact the average and standard deviation of <D/H> = 28.6±8.2 ppm implies that some astrophysical effect is producing different values of D/H in the relatively low metallicity IGM absorbers or the observationally assigned errors are about a factor of two too small. If the assigned observational errors are too small, the error in the average will be smaller than the standard deviation by about a factor of $5^{0.5} = 2.2$. However, although such an error calculation makes the average appear to be accurately determined, it is premature to reduce the error in the average by such a large factor without a better understanding of the observational errors in the IGM D/H measurements. We therefore list an average IGM value of D/H in Table 6 along with the standard deviation as the error.

*Primordial D/H*. The best estimate of the primordial value of D/H is derived from measurements of the baryon/photon ratio from the fluctuations in the cosmic microwave background radiation coupled with the theory of element production in the big bang. We use the most recent Wilkinson Microwave Anisotropy Probe (WMAP) observations of the baryon/photon ratio based on the three year WMAP data of Spergel et al. (2006). Using the Steigman (2005) conversion between the baryon/photon ratio and D/H in the standard $\Lambda$CDM universe we obtain D/H = 25.9±1.2. The ±1$\sigma$ range allowed by these WMAP observations is shown in Figure 6 as the dashed lines on the right hand side of the figure. Slightly different values of D/H are obtained when using different nuclear reaction rates to describe the big bang nucleosynthesis. For example, Cyburt et al. (2003) obtain D/H = 27.5(+2.4, -1.9) ppm while Coc et al. (2004) obtain D/H = 26.0 (+1.9, -1.7) using the earlier WMAP results of Spergel et al. (2003) which are nearly identical to those of Spergel et al. (2006) in the case of the baryon/photon ratio.

### 7.2. A Comparison of the Observed Values of D/H
### in the Different Astrophysical Sites

Figure 6 summarizes the measured values of D/H in the different astrophysical sites discussed in §7.1 and tabulated in Table 6. Selective depletion of D into dust grains may be affecting some of the measured values of D/H (Draine 2004, 2006; Linsky et al. 2006). In the third column of Table 6 we indicate the possible importance of dust depletion effects for the different astrophysical sites. Discussions of our assignments as not important, important, or probably important are given below. We have plotted those data points whose values of D/H are unlikely to be affected by dust depletion with filled symbols and those data points with values of D/H possibly affected by dust depletion with half-filled and open symbols.

The D/H data points unlikely to be affected by the incorporation of D into dust include the D/H values for the IGM, for Complex C, for the WNM of the Galactic halo, for the high values of D/H for the Solar neighbohood, and possibly for Jupiter. The IGM values of D/H are for low metallicity environments that likely have very little dust. There is no evidence for dust in Complex C and no $H_2$ has been detected (Richter et al. 2001; Sembach et al. 2004; Wakker 2006). The gas in the WNM of the Galactic halo contains dust (Sembach & Savage 1996 ) but the energetic events that lead to the expulsion of the gas and dust from the disk into the halo likely destroys much of the dust except for the refractory cores of the grains (Sembach & Savage 1996 ). This result follows from studies of the elemental abundances in the WNM found at different distances away from the Galactic plane. The high values of D/H measured for 5 Solar neighborhood lines of sight have been proposed by Linsky et al. (2006) to likely represent lines of sight where the effects of D depletion into dust are minimal.

In the case of Jupiter, we assume that the gaseous material in the protostellar disk that became incorporated into Jupiter contained dust but that the deuterium possibly incorporated into that dust became liberated into the gas phase as Jupiter was gravitationally assembled. However, a 5-10% enrichment of D in the atmosphere of Jupiter may have occurred if Jupiter was formed from deuterium enriched ices in addition to normal protostellar disk material (Guillot 1999). Thus, the value of D/H in the atmosphere of Jupiter (possibly reduced by 5-10%) should be a fair measure of the value of D/H in the ISM out of which the Sun and Jupiter formed 4.6 Gyr ago provided there has not been a subsequent separation of HD and $H_2$ in the different gaseous and liquid phases of $H_2$ in Jupiter.

The astrophysical sites where the effects of the incorporation of D into dust are expected to be largest include the distant Galactic disk and the 12 lines of sight into the Solar neighborhood where low values of D/H are observed ( see Table 6).  The Local Bubble may represent an intermediate situation where modest levels of D depletion are present  (Linsky et al. 2006).  However, mixing of interstellar gas  with different intrinsic abundances of D may also play a role in regulating the average D/H ratio locally.

The new result for the WNM of the Galactic halo, D/H = 22 (+8, -6),  is consistent with the Linsky et al.  (2006) explanation for the variations seen in the measured values of D/H in the Galaxy.  The WNM of the Galactic halo is believed to be gas that originates in the thin disk and is elevated into the halo and supported through energetic events in the disk and low halo.  Those energetic events  (supernova explosions) result in partial grain destruction so the dust in halo gas is believed to mostly only contain the highly refractory cores of grains.   The similarity of the values of D/H in the halo with those recorded for the 5 solar neighborhood sight lines exhibiting high D/H is consistent with the  Linsky et al. (2006) suggestion that lower D/H values observed in the disk occur in regions affected by dust depletion effects.

It is possible the value of D/H in the WNM of the halo is affected by the infall of D rich gas from beyond the Milky Way.  The  low metallicity Complex C  is evidence for such an infall (Wakker et al. 1999).  However, while the effect of infall would be to increase the value of D/H in the WNM of the halo, it would also decrease the value of O/H in the halo gas.  In the case of the line sight to HE 0226-4115 O/H appears to be roughly Solar in abundance.  It therefore appears that infall is not substantially modifying the abundances along this line of sight.

Our result for the WNM of the Galactic halo essentially requires that the low values of D/H found in the Galactic disk and in the Solar neighborhood and possibly even in the Local Bubble are low because a substantial fraction of  the D is somehow hidden from detection  (i.e. some of the D is incorporated into dust).  If that were not true,  it would be difficult to understand how matter flowing from the disk  with D/H ~ 10 ppm increases to ~ 22 ppm after arriving in the halo.  Destruction of dust containing D provides a natural explanation.

### 7.3.  *Implications for Galactic Chemical Evolution*

If the depletion explanation for the low values of D/H in Figure 6 is correct, the only values of D/H in the Galaxy that are useful  to probe Galactic chemical evolution are the average for the 5 high values for the Solar neighborhood  [<D/H> = 23.1±2.4 (1$\sigma$) ppm] , the value for the WNM in the halo [D/H = 22 (+8, −6) ppm],  and possibly the value for Jupiter [23.0±2.6 (1$\sigma$) ppm].   The Solar neighborhood and halo results listed here have both been corrected for foreground contamination by H I and D I absorption in the Local Bubble. We note that Linsky et al. (2006) reported their Solar neighborhood average value of D/H as a lower limit since they were concerned that a small amount of dust depletion may still be affecting the result.  While that is a possible problem, an effect that would tend to cause the true average to be a smaller value is related to the fact that the Linsky et al. 5 star sample simply represents an average over the 5 stars with the largest values of D/H in their 29 star sample.  It is possible that some of those data points yield large values of D/H because of occasional positive 2$\sigma$ statistical fluctuations in the measured values of D/H for the larger sample.

The values of D/H above for Jupiter and for the 5 star average are only slightly smaller than the assumed primordial value inferred from the WMAP measurements [D/H = 25.9±1.2 (1σ)]. This implies the level of D astration in the Galaxy has been small over 9 and 14 Gyrs since its formation.

The value of <D/H> = 23.0±2.6 for Jupiter implies that the abundance of D in the Galaxy only changed by a factor $f_d$= (25.9±1.2)/ 23.0±2.6) = 1.13±0.14 over the first ~8 Gyrs of its evolution. If we reduce the Jovian D/H measurement by 7% to allow for the possibility that Jupiter formed from normal protostellar disk matter and deuterium enriched ices (Guillot 1999) the astration factor increases to $f_d$ = 1.20.

The high value of <D/H> for the five solar neighborhood stars (corrected for Local Bubble contamination) from Linsky et al. (2006) which seem to be unaffected by depletion implies the abundance of D changed by a factor $f_d$ = (25.9±1.2)/(23.1±2.4) = 1.12±0.13 over ~13 Gyrs of evolution of the Milky Way. If dust containing D still exists in appreciable quantities along some of the five sight lines, the actual astration factor will be smaller.

Models of Galactic chemical evolution currently predict larger values of the D astration factor. For example, the models of Chiappini, Renda, & Matteucci (2002) predict $f_d$ ~1.5 while models of Romano et al. (2006) can accommodate values of $f_d$ ranging from 1.83 to 1.39. The theoretical deuterium astration factor is sensitive to numerous assumptions including properties of the initial mass function, stellar lifetimes, stellar mixing, and the rate of infall of unprocessed matter. There may be enough uncertainty associated with these and other galactic chemical evolution assumptions to produce the smaller observed astration factor of 1.12 without violating other abundance constraints.

We thank the HST and FUSE mission operations teams for their efforts to provide excellent UV and FUV spectroscopic data to the astronomical community. We thank Marilyn Meade for assistance with the FUSE data handling at the University of Wisconsin. Helpful comments about a draft version of this manuscript were provided by Jeffrey Linsky. We thank the anonymous referee for a number of helpful suggestions regarding how the submitted manuscript could be improved. The STIS observations of HE 0226-4110 were obtained for HST program 9148 with financial support to B.D.S. through NASA grant HST-GO-9184.08-A from the Space Telescope Science Institute (STScI). B.D.S. also acknowledges support for the FUSE observations and interpretations through NASA grant NNG-04GC70G. The data obtained in this paper were obtained from the Multi-mission Archive at STScI (MAST).

TABLE 1
ABSORPTION LINE RESULTS

| Ion | λ (Å) | Log(λf) | $W_\lambda{}^a$ (mÅ) | log$W_\lambda/\lambda$ | Day/ Night | Channel | Note |
|---|---|---|---|---|---|---|---|
| (1) | (2) | (3) | (4) | (5) | (6) | (7) | (8) |
| D I | 949.485 | 1.122 | 72.0±18.2 | -4.12 | N | SiC2A | 1 |
| D I | 937.548 | 0.864 | 65.3±18.7 | -4.16 | N | SiC2A | 1 |
| D I | 930.495 | 0.652 | 49.4±13.2 | -4.27 | N | SiC2A | 1 |
| D I | 925.974 | 0.470 | 41.6±17.3 | -4.35 | N | SiC2A | 1 |
| D I | 922.899 | 0.311 | 19.2±13.3 | -4.68: | N | SiC2A | 1 |
| D I | 920.716 | 0.170 | 13.5±15.2 | -4.83: | N | SiC2A | 1 |
| D I | 919.101 | 0.043 | 13.3±19.5 | -4.84: | N | SiC2A | 1 |
| O I | 1302.169 | 1.796 | 197.8±10.7 | -3.82 | N+D | STIS | |
| O I | 1039.230 | 0.974 | 108.1±11.1 | -3.98 | N | LiF1A | 2 |
| O I | 976.448 | 0.509 | 115.8±22.4 | -3.93 | N | SiC2A | |
| O I | 974.070 | -1.817 | <36.8 (3σ) | <-4.42 | N+D | SiC2A | |
| O I | 971.738 | 1.052 | 89.1±15.2 | -4.04 | N | SiC2A | 2 |
| O I | 950.885 | 0.176 | ... | ... | N | SiC2A | 3 |
| O I | 948.686 | 0.778 | 97.5±19.3 | -3.99 | N | SiC2A | 2 |
| O I | 937.841 | -0.085 | ... | ... | N | SiC2A | 3 |
| O I | 936.630 | 0.534 | 85.3±15.2 | -4.04 | N | SiC2A | 2 |
| O I | 930.257 | -0.301 | 80.1±17.8 | -4.06 | N | SiC2A | |
| O I | 929.517 | 0.329 | 97.8±14.8 | -3.98 | N | SiC2A | |
| O I | 925.446 | -0.484 | 77.6±12.7 | -4.08 | N | SiC2A | |
| O I | 924.950 | 0.155 | 105.1±16.2 | -3.94 | N | SiC2A | 3 |
| O I | 922.200 | -0.645 | 58.8±17.1 | -4.20 | N+D | SiC2A | |
| O I | 921.857 | -0.001 | 89.3±18.1 | -4.01 | N | SiC2A | |
| O I | 919.917 | -0.788 | ... | ... | N | SiC2A | 3 |
| Mg II | 1239.925 | -0.106 | <25.7 (3σ) | <-4.68 | N+D | STIS | |
| Mg II | 1240.395 | -0.355 | <26.9 (3σ) | <-4.66 | N+D | STIS | |
| P II | 1152.818 | 2.451 | 57.3±10.2 | -4.30 | N+D | LiF2A | |
| S II | 1259.518 | 1.320 | 121.5±8.1 | -4.02 | N+D | STIS | |
| S II | 1253.805 | 1.136 | 101.3±9.5 | -4.09 | N+D | STIS | |
| S II | 1250.578 | 0.832 | 72.1±11.1 | -4.24 | N+D | STIS | |
| S III | 1190.203 | 1.449 | 79.0±15.8 | -4.18 | N+D | STIS | |
| Fe II | 1608.451 | 1.968 | 202.6±21.2 | -3.90 | N+D | STIS | |
| Fe II | 1144.938 | 1.978 | 123.1±10.6 | -3.97 | N+D | LiF2A | |
| Fe II | 1143.226 | 1.306 | 71.5±8.2 | -4.20 | N+D | LiF2A | |
| Fe II | 1142.366 | 0.661 | 32.3±8.5 | -4.55 | N+D | LiF2A | |
| Fe II | 1127.098 | 0.102 | <24.9 (3σ) | <-4.66 | N+D | LiF2A | |
| Fe II | 1125.448 | 1.244 | 68.2±9.1 | -4.22 | N+D | LiF2A | |
| Fe II | 1121.975 | 1.512 | 82.7±8.2 | -4.13 | N+D | LiF2A | |
| Fe II | 1112.048 | 0.695 | 26.6±7.3 | -4.62 | N+D | LiF2A | |
| Fe II | 1096.877 | 1.554 | 84.1±11.1 | -4.12 | N+D | LiF2A | |
| Fe II | 1063.176 | 1.765 | 104.4±7.1 | -4.01 | N+D | LiF1A | |
| Fe II | 1062.152 | 0.490 | 14.9±4.6 | -4.85 | N+D | LiF1A | |
| Fe II | 1055.262 | 0.812 | 28.9±6.8 | -4.56 | N+D | LiF1A | |
| Ni II | 1370.132 | 1.906 | 43.2±9.7 | -4.50 | N+D | STIS | 4 |
| Ni II | 1317.217 | 1.876 | 84.3±11.0 | -4.19 | N+D | STIS | 4 |

$^a$ Errors are ±1σ. Limits are 3σ detection limits.
Notes.- (1) See § 5.4 for a discussion of the DI equivalent
width measurements and their errors. (2) Airglow strong, line
not used in the PF or GOG analysis. (3) Blended with another absorption line, not used in the analysis. (4) The two Ni II lines should have similar equivalent widths given the two values of log ($\lambda$f) are similar. This implies that either one of the measurements is affected by emission or the other is affected by blended absorption. While Ni II $\lambda$1317 is partially blended with H I Ly $\alpha$ at z = 0.08372 (see Lehner et al. 2006) what appeared to be appropriate deblending resulted in the listed equivalent width.

TABLE 2
COLUMN DENSITIES AND LINE WIDTHS FOR THE
PRIMARY ABSORPTION COMPONENT NEAR –4 km s$^{-1}$

| Ion | Method | $v_{LSR}$ (km s$^{-1}$) | b (km s$^{-1}$) | LogN(ion) | Note |
|---|---|---|---|---|---|
| S II | COG | -4.1 | 10.8(+2.8,-1.9) | 15.16(+0.18.-0.12) | |
| S II | PF | -4.1±0.3 | 10.0±0.5 | 15.31±0.04 | 1 |
| S II | PF | -1.2±0.9 | 10.1±1.5 | 15.19±0.04 | 2 |
| O I | COG | -3.6 | 8.3±0.7 | 16.95(+0.37,-0.20) | 3 |
| O I | PF | -3.7±0.3 | 8.5±0.3 | 16.86±0.10 | 3 |
| Fe II | COG | -3.5 | 10.5±0.9 | 14.77±0.06 | |
| Fe II | PF | -3.5±0.3 | 11.1±0.4 | 14.83±0.03 | |
| P II | AOD | -4.0 | … | 13.37±0.07 | |
| S III | PF | -4.1±2.6 | 14.8±6.6 | 14.56±0.09 | |
| D I | COG | -7.5 | 8.6(+5.9,-2.4) | 15.33(+0.24,-0.17) | 4 |
| D I | PF | -7.5±1.3 | 10.4±2.1 | 15.45±0.08 (+0.04,-0.06) | 5 |
| H I | PF | -4.1 | 10.7 | 20.12±0.02 (+0.00,-0.04) | 6 |

Notes.- (1) Results are for a simultaneous PF to all three S II lines. (2) Results are for a PF to the weakest S II $\lambda$1250 line. (3) The O I lines used for the COG and PF analysis are $\lambda\lambda$ 1302, 976, 930, 929, 925, 924, 922, and 921. For the PF the non-detected O I $\lambda$974 line is also included. Omitted O I lines are blended or strongly affected by O I airglow emission (see Table 1). (4) COG fit to the D I $\lambda\lambda$949, 937, 930, 925, 922, 920, 919 lines. (5) Simultaneous PF applied to the D I $\lambda\lambda$949, 937, 930, 925, 922, 920, 919 lines with FWHM of the instrumental LSF = 25 km s$^{-1}$. The first error is the PF random error. The second error is the combined systematic errors associated with continuum placement uncertainty and the uncertainty in the instrumental profile width (see §5.5). (6) The H I PF methods and results are discussed in §5.5. The first error is the random PF fit error. The second error is the systematic continuum placement error.

TABLE 3
APPROXIMATE COLUMN DENSITIES FOR METALS IN IVC1 AND IVC2[a]

| Ion | Method | $v_-$ (km s$^{-1}$) | $v_+$ (km s$^{-1}$) | $v_{LSR}$ (km s$^{-1}$) | b (km s$^{-1}$) | LogN(ion) | Note |
|---|---|---|---|---|---|---|---|
| O I | PF | … | … | -33.6: | 7.5 | 13.06±0.32 | 1 |
| O I | AOD | -43 | -28 | -33.3±1.6 | 4.2±1.7 | 13.22(+0.12, -0.18) | 2 |
| C II | AOD | -70 | -33.8 | -43.2 | … | 13.41±0.07 | 3 |
| C III | AOD | -70 | -33.8 | -45.1 | … | >13.19(+0.13, -0.17) | 6 |
| Si II | AOD | -70 | -33.8 | -46.2 | … | 12.35±0.09 | 4 |
| Si III | PF | … | … | -47.9±3.3 | 5.21: | 12.02±0.27 | 5 |

[a] $v_-$ and $v_+$ indicate the velocity range of the AOD integrations. $v_{LSR}$ is the average velocity of the absorption derived from the PF or AOD integration.
  Notes.- (1) PF to the weak IVC2 seen in the O I λ1302.17 line. The component b value was fixed at 7.5 km s$^{-1}$ to constrain the fit. (2) AOD results for IVC2 seen in the O I λ1302.17 line. (3) Based on AOD measurements of IVC1 seen in the C II λ1334.53 line. (4) Based on AOD measurements of IVC1 seen in the Si II λ1260.42 line. (5) Based on PF measurements of IVC1 seen in the Si III λ1206.50 line. (6) The C III λ977 IVC absorption is strong. The AOD value of the column density should be considered a lower limit because of the possibility of unresolved saturation.

TABLE 4
PROPERTIES OF THE H I VELOCITY COMPONENTS[a]

| Component | $v_{LSR}$ (km s$^{-1}$) | b (km s$^{-1}$) | logN(H I) | Note |
|---|---|---|---|---|
| IVC2 | -33.6 | 14 | 16.67 | 1 |
| Primary | -4.1 | 10.7 | 20.12±0.02(+0.00,-0.04) | 2 |
| HVC1 | 92±5 | 25±2 | 16.29±0.05 | 3 |
| HVC2 | 146±7 | 15±2 | 16.21±0.10 | 3 |
| HVC3 | 173±5 | 12±1 | 16.74±0.20 | 3 |
| HVC4 | 200±6 | 18±3 | 16.34±0.10 | 3 |

[a] We have not attempted to model the weak IVC3 absorption in the +20 to +40 km s$^{-1}$ velocity range that is evident in the strong C II and Si III absorption lines. This absorption has no appreciable influence on the derived values of N(D I) or N(H I).
  Note.- (1) Component IVC2 has a velocity established by the weak O I absorption component seen in the O I λ1302.17 profile. The value of logN(H I) is scaled from logN(O I) = 13.22 assuming log(O/H)$_{ISM}$ = -3.45. The b value was determined to give the best fit to the wing of the higher Lyman series absorption lines over the velocity range from -30 to -60 km s$^{-1}$(see right panel of Fig. 3). This absorption component also likely includes blended contributions from H I absorption associated with IVC1. (2) Most of the H I along the line of sight to HE 0226-4110 is in the primary component. The adopted velocity is taken to be the velocity of the S II absorption. The b value is taken to be that determined from the high resolution HI 21 cm emission observations. (3) The properties of these high velocity components are taken from the study of the HVCs toward HE 0226-4110 by Fox et al. (2005).

TABLE 5
ADOPTED COLUMN DENSITIES FOR THE
PRIMARY ABSORPTION COMPONENT AT −4 km s$^{-1}$

| Ion | logN(ion)±1σ | log[N(X)/N(HI)] | Note |
|------|--------------|------------------|------|
| H I | 20.12±0.02(+0.00,-0.04) | 0.00 | 1 |
| D I | 15.45±0.08 (+0.04, -0.06) | -4.67 ±0.14 | 1 |
| O I | 16.90 (+0.37,-0.20) | -3.22(+0.41,-0.20) | 2 |
| P II | 13.37±0.07 | -6.75(+0.11,-0.07) | 3 |
| S II | 15.25±0.10 | -4.87(+0.14,-0.10) | 4 |
| S III | 14.56±0.09 | -5.56(+0.13, -0.09) | 1 |
| Fe II | 14.80(0.06,-0.09) | -5.32(+0.10,-0.09) | 5 |

Notes.-
 (1) The PF values of the column density and errors. For  D I and H I the first error is the PF statistical error and the second error is the systematic  error (see §5.5 for H I and §5.4 for D I).
(2) We average the COG and PF values of logN(O I) but adopt the larger COG errors because the weakest O I lines are highly saturated and the derived value the column density is very sensitive to the validity of the simple assumed kinematic model of one absorbing component.
 (3) The AOD value of the column density and error are adopted.
(4) The two PF values of logN(S II) listed in Table 2 are averaged and the adopted error spans the full ±1σ range of the two PF results.
(5) The PF and COG values of logN(Fe II) are averaged and the 1σ error is adopted to span the full range of values of logN(Fe II)±1σ from both methods.

TABLE 6
SITES OF D/H MEASUREMENTS

| Site | D/H (ppm) | Possible Importance of dust Depletion | Number of measurements | Sources |
|---|---|---|---|---|
| Primordial gas | 25.9±1.2 (1σ) | Not important | 1 | 1 |
| IGM | 28.6±8.2 (SD) | Not important | 6 | 2 |
| Complex C | 22±7 (1σ) | Not important | 1 | 3 |
| Galactic halo WNM | 22 ±(+8,-6) (1σ) | Not important | 1 | 4 |
| Galactic disk | 8.5±1.0 (SD) | Important | 4 | 5 |
| Solar Neighborhood (high) | 23.1±2.4 (1σ) | Probably not important | 5 | 6 |
| Solar Neighborhood (low) | 10.1±3.3(SD) | Important | 12 | 7 |
| Local Bubble | 15.8±2.1 (SD) | Probably important | 22 | 8 |
| Jupiter | 23.0±2.6(1σ) | Probably not important | 3 | 9 |

Source:   (1) The primordial gas value and ±1σ errors are based on the recent Spergel et al. (2006) WMAP measure of the baryon to photon ratio combined with the Steigman (2005) conversion between D/H and the baryon to photon ratio in the standard lambda-CDM universe.  (2) The 6 individual measurements of D/H in the IGM shown in Fig. 6 are from the summary of reliable values from Pettini (2006) supplemented by the recent O'Meara et al. (2006) result for SDSS 1558-0031. These measurements yield the average and standard deviation listed here.  The standard deviation is much larger than the ±1σ error bars attached to each measurement.  (3) The value and 1σ error for a line of sight through HVC Complex C is from Sembach et al. (2004). (4) The result and 1σ (random+systematic) error for the WMN of the lower Galactic halo is from this paper. The listed value includes a +1 ppm correction for the presence of foreground D I and H I in the Local Bubble. (5) We list the average and standard deviation for D/H  for the four Galactic disk lines of sight in the Linsky et al. (2006) compilation with logN(H I) >20.7. (6) We list the mean value  of D/H for the 5 high values of D/H observed in the solar neighborhood with logN(HI) ranging from 19.2 to 20.6 as compiled  by Linsky et al. (2006).  The value listed allows  for the Local Bubble foreground contamination of these observations (see §7.1). To minimize potential systematic errors,  Linsky et al. (2006) derived a straight mean rather than an error weighted mean.  (7) We list the average and standard deviation for D/H for the 12 low values of D/H observed in the Solar neighborhood as compiled by Linsky et al. (2006).   (8) We list the average and standard deviation for D/H for the 22 values of D/H compiled by Linsky et al. (2006) for the Local Bubble with logN(HI) ranging from 17.6 to 19.1.  (9) The result listed for Jupiter is an error weighted average of the measurements from Trauger et al. (1973),  Mahaffy et al. (1988) and Lellouch et al. (2001). The ±1σ error in the average was obtained from the normal rules of combining measurements and their errors.

FIGURES

FIG. 1. Metal line absorption profiles from FUSE and STIS observations including profile fits to selected lines. The continuum normalized flux is plotted against $v_{LSR}$ with the ion identified in the lower left of each panel. STIS observations for lines with $\lambda >$ 1180Å are binned to 3.5 km s$^{-1}$. FUSE observations for lines with $\lambda <$ 1180Å are binned to 8 km s$^{-1}$. The vertical dashed lines denote the velocity of the principal H I absorption component. The vertical dotted lines for C II, C III and Si III denote the AOD integration range for the negative velocity IVC absorption.

FIG. 2. – COGs are shown for absorption produced by O I, S II, Fe II and D I in the spectrum of HE 0226-4110. The wavelengths of each line used in the final COG fits are listed next to the value of $\log W_{\lambda}/\lambda$. The final fit parameters are shown in each panel and listed in Table 2. The dimensions of $Nf\lambda$ are cm$^{-1}$.

FIG. 3. – D I and H I absorption profiles and H I ISM absorption models for the line of sight to HE 0226-4110. In the left panel, the raw night-only FUSE observations are shown with the velocity referenced to the H I Lyman series absorption. H I airglow emission contaminates the H I Ly 972, 949, 937, and 930 absorption. An empirical fit to the airglow emission (see the red lines on the left panel) has been subtracted from the observations to give the corrected profiles displayed in the right panels. In both sets of panels the D I absorption is highlighted in yellow. In all panels the observations are binned to four pixels corresponding to 8 km s$^{-1}$. The continuum and zero levels are shown with the dashed lines. In the right panels the blue line shows the H I absorption model without the weak component at –33.6 km s$^{-1}$. The red line shows the model including a component at –33.6 km s$^{-1}$ with b = 14 km s$^{-1}$ and logN(H I) = 16.67. The D I absorption was analyzed using the red line H I absorption model for the continuum placement on the positive velocity side of the D I absorption. The H I absorption model assumes FWHM = 25 km s$^{-1}$ for the FUSE line spread function.

FIG. 4. Profile fits to the absorption in the lines of D I $\lambda\lambda$949, 930, 925, 922 , 920 and 919. The histogram shows the FUSE observations from Fig. 3 displayed in the D I reference frame. The D I observations shown are not affected by H I airglow contamination. The upper solid line is the H I model shown in the right panel of Fig. 3 in red. The lower solid line shows the D I fit result for logN(D I) = 15.45 and b = 10.4 km s$^{-1}$. The H I and D I model absorption assume FWHM = 25 km s$^{-1}$ for the FUSE line spread function.

FIG. 5. – HI absorption line profiles toward HE 0226-4110 showing the derivation of the H I column density using a fit to the strong damping wings to Ly $\alpha$. In the top panel we display the STIS Ly $\alpha$ profile over a broad velocity range to illustrate the large scale continuum behavior. In the second and third panels from the top we display the same observations over a smaller velocity range with 3.5 km s$^{-1}$ pixels in the second panel and 52.4 km s$^{-1}$ pixels in the third panel. In all three panels for the Ly $\alpha$ line a first order polynomial fit to the continuum is shown as the nearly horizontal line in green. The profile fit to the Ly α absorption is displayed in blue in the second and third panels, with the three curves in the third panel showing logN(H I) = 20.12±0.04 (2σ) where the ±2σ error does not include the systematic uncertainties associated with the continuum placement. The bottom panel shows the FUSE observations of Ly β. The fitted blue curve is for logN(H I)= 20.12 in the primary component and the H I IVC and HVC column densities listed in Table 4. H I geocoronal emission is visible in all four panels. The Ly β panel in addition shows O I airglow emission near 400 and 700 km s⁻¹.

FIG. 6. – Values of D/H measured for different sites including: Jupiter, the Local Bubble, the Solar neighborhood (high and low values), the more distant galactic disk, the lower Galactic halo (this paper), Complex C, and the IGM (see Table 6). The primordial value shown with the dashed lines is inferred from WMAP observations of the baryon to photon ratio combined with the predictions of big bang nucleosynthesis in a lambda-CDM universe. Numbers listed for each absorption site indicates the number of individual observations contributing to the average value for the plotted data point. The errors shown are usually the dispersion about the average. For the IGM the individual high quality data points are shown. Note that the dispersion in those points is much larger than the indicated 1σ observational errors. See Table 6 for the sources of the plotted observations.

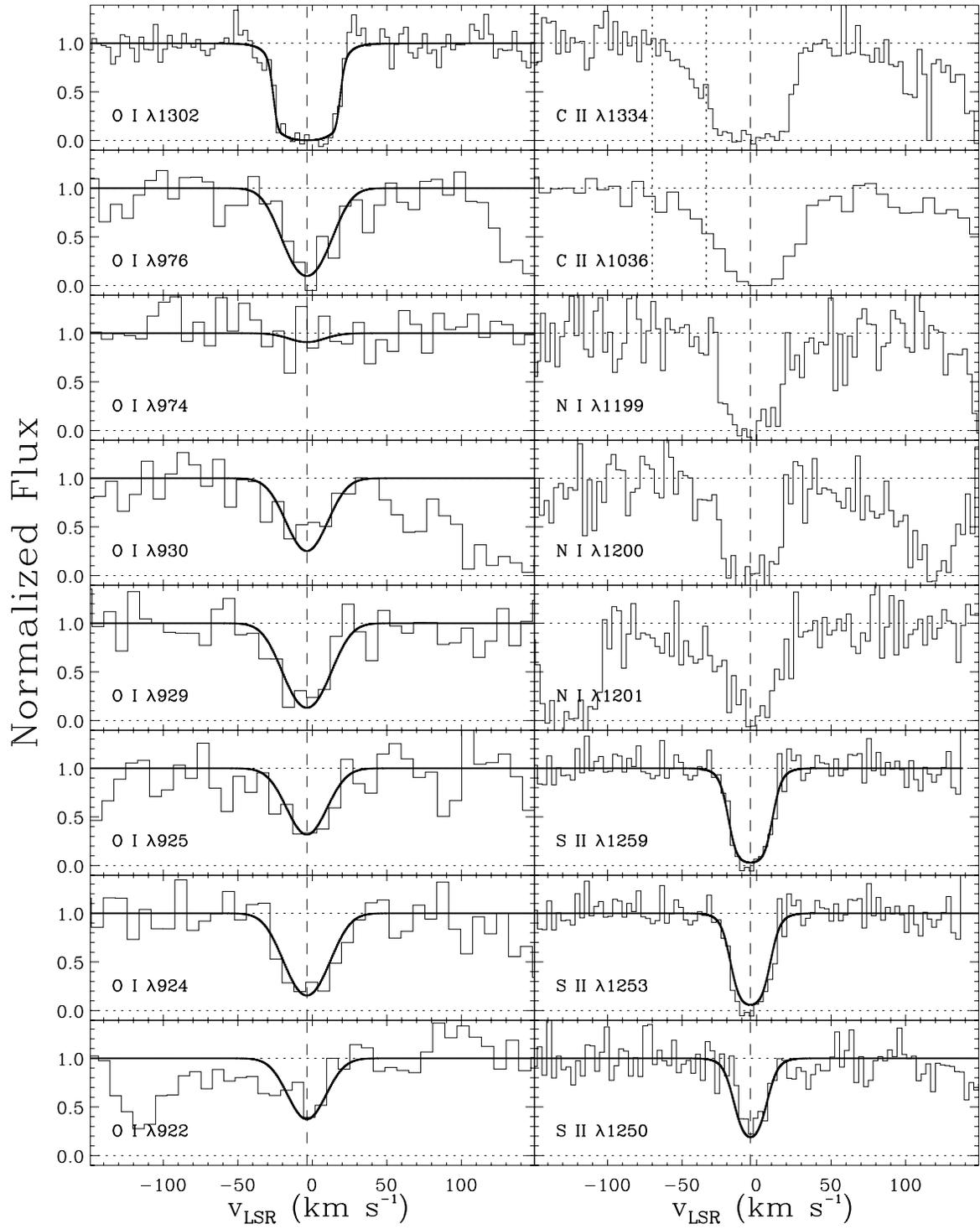

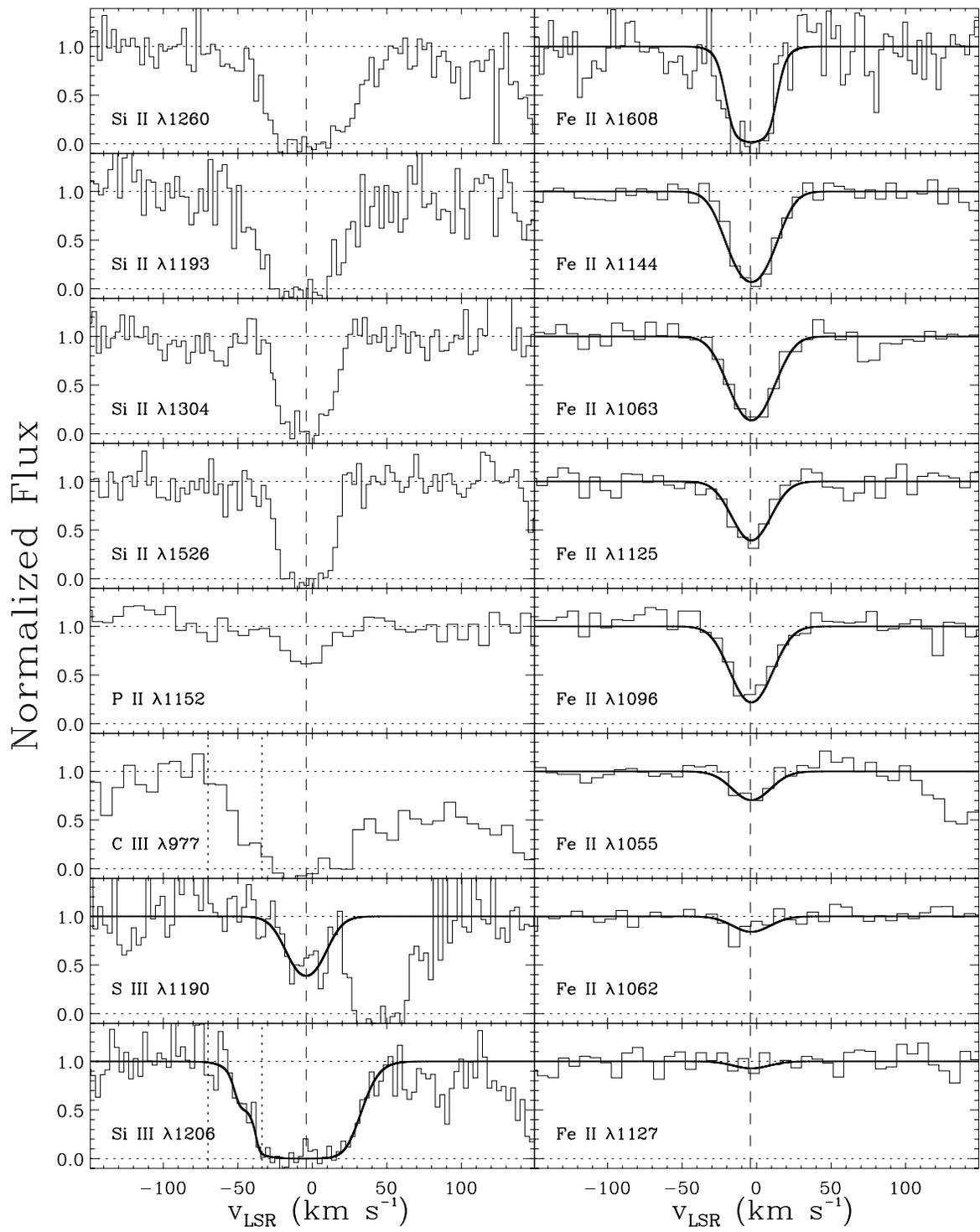

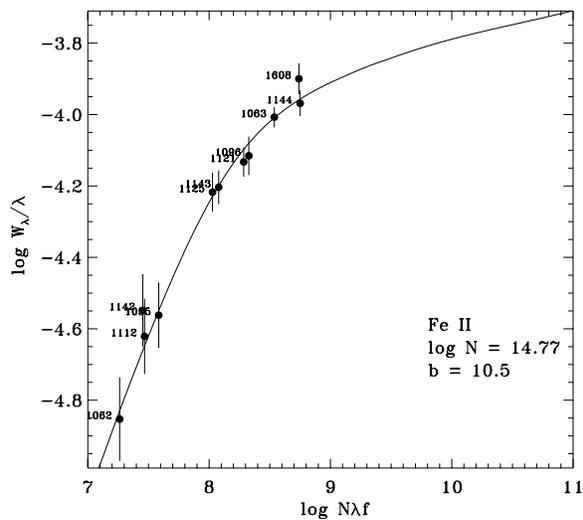
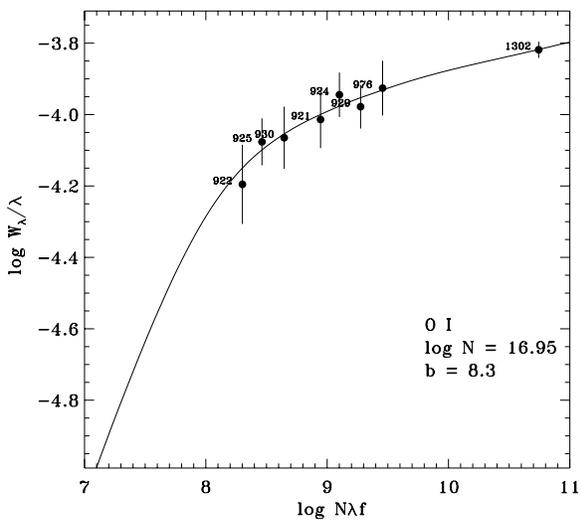
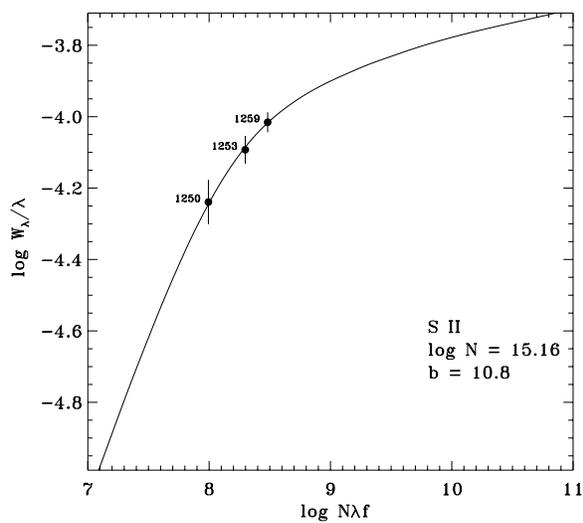
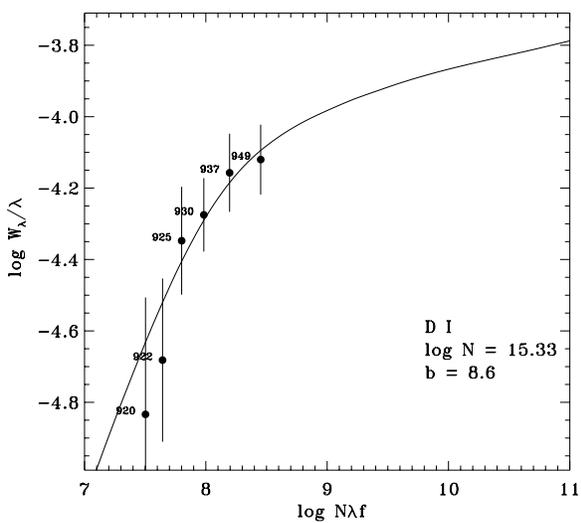

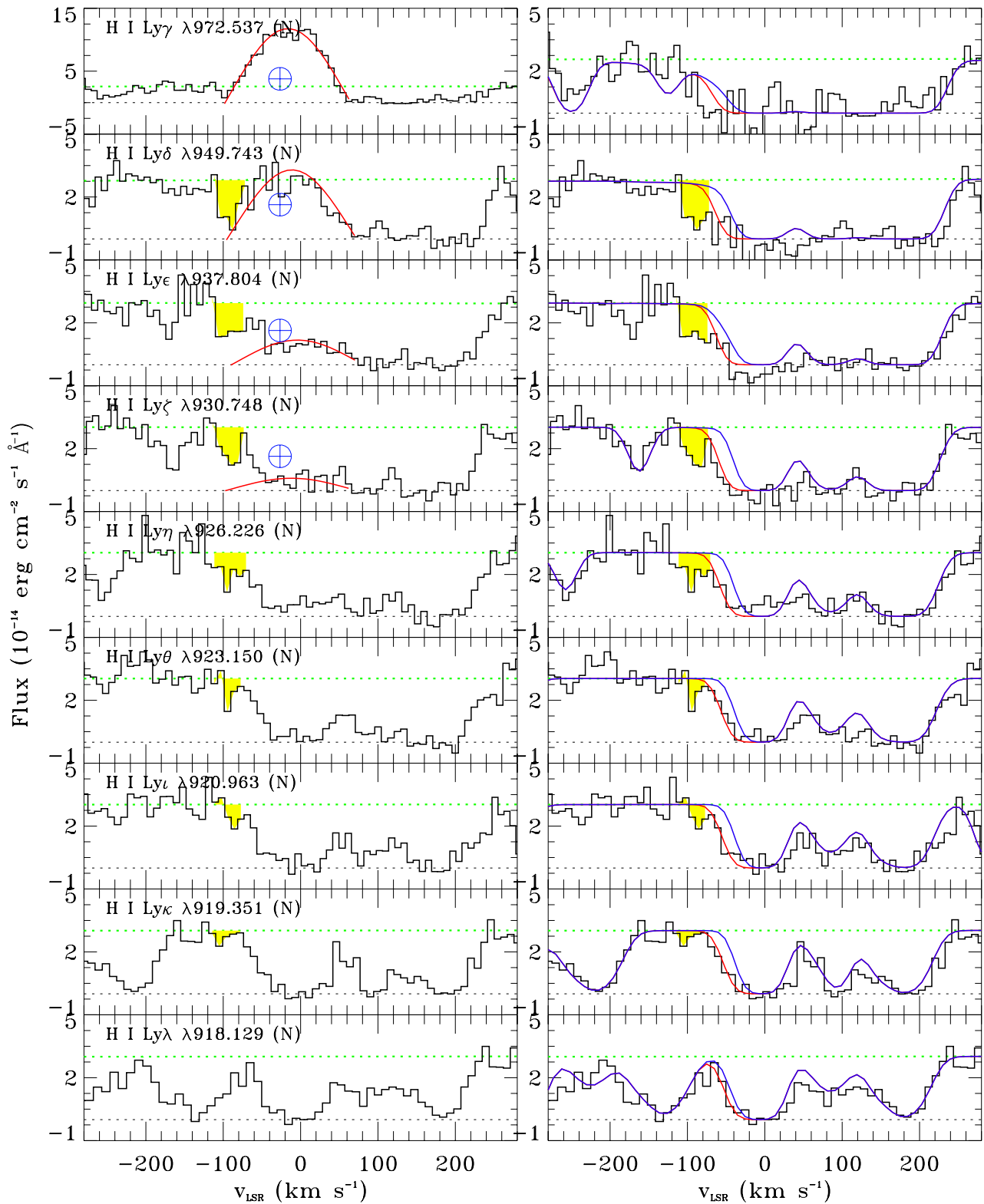

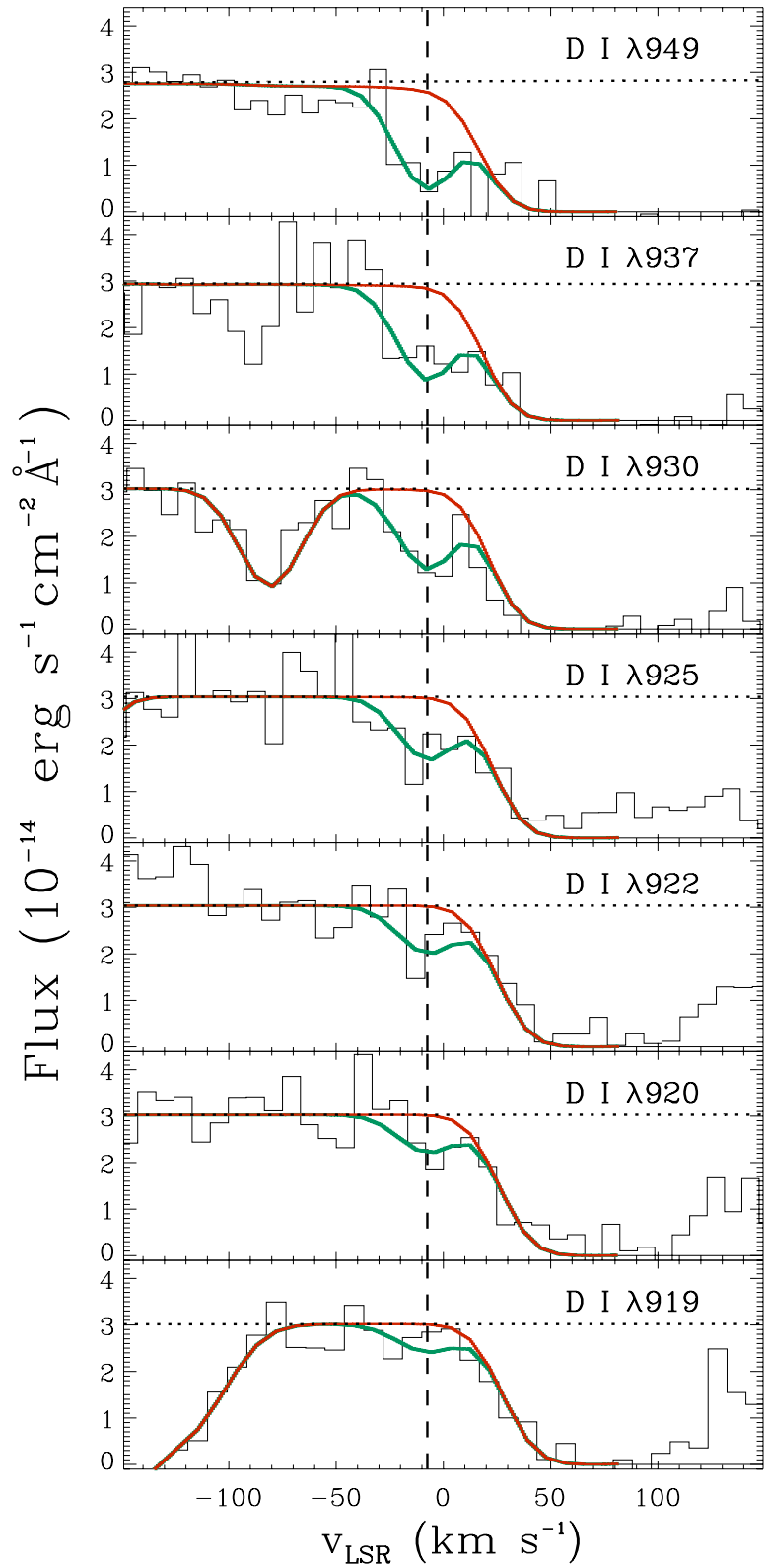

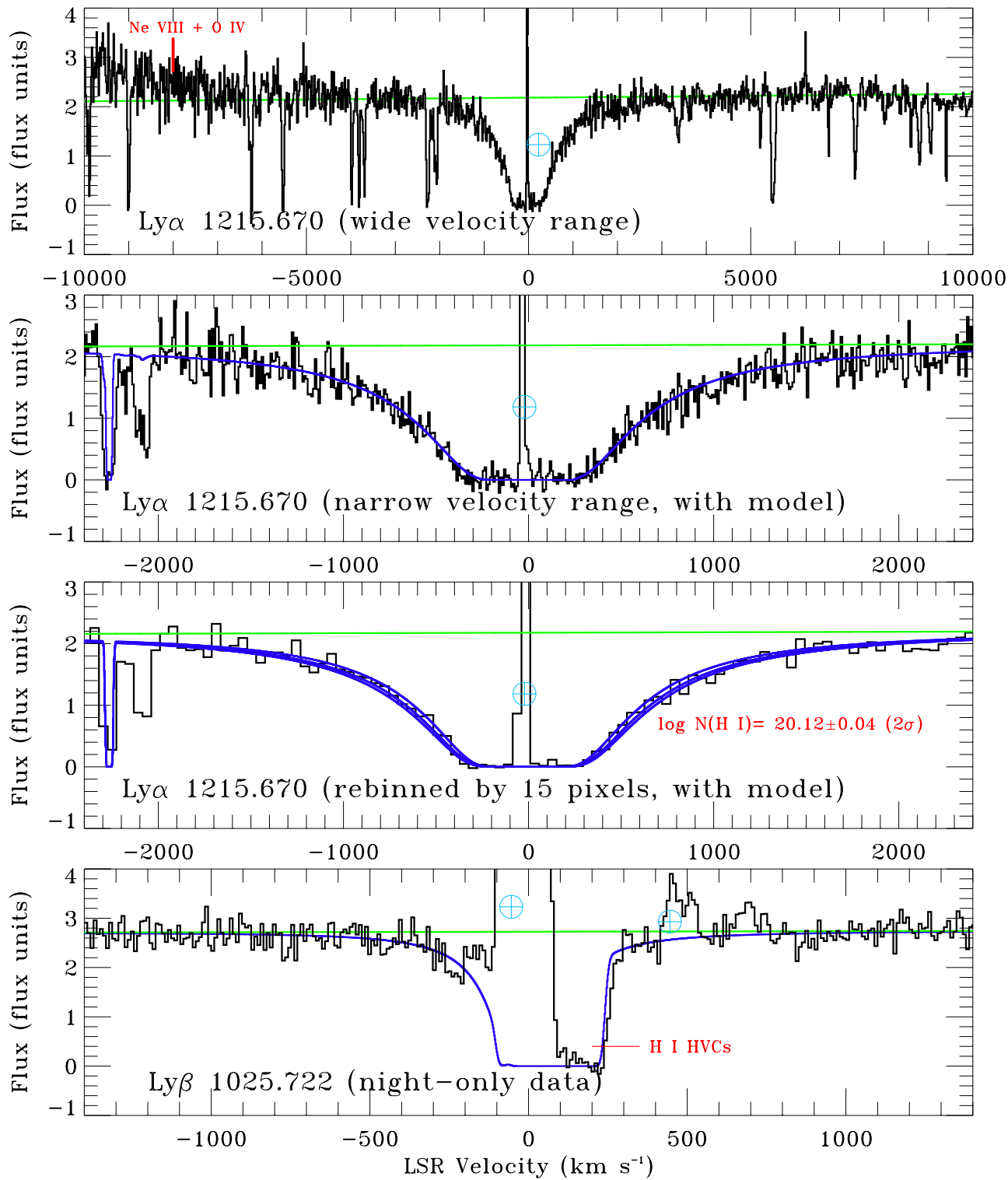

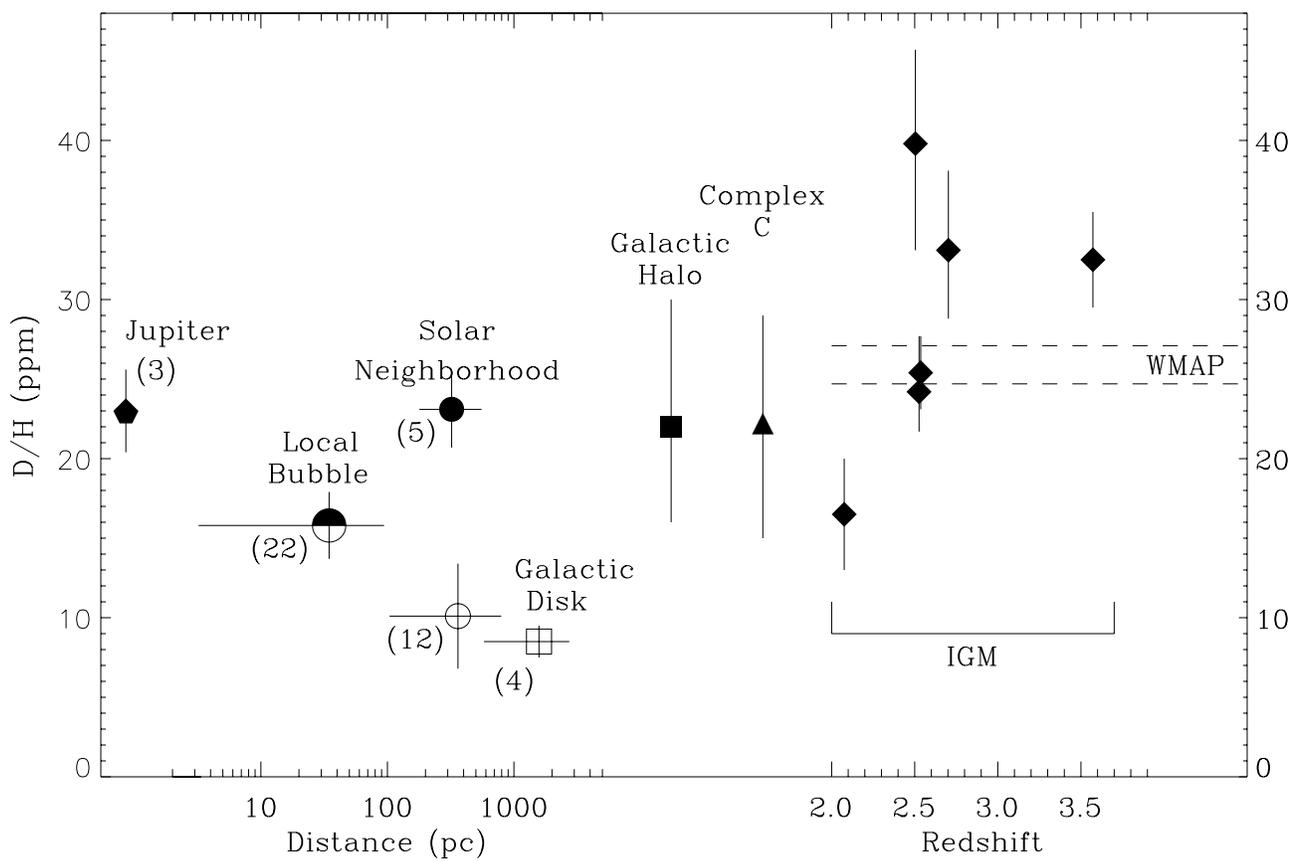